\documentclass[preprint,aps,superscriptaddress]{revtex4}

\usepackage{amsmath}
\usepackage{amssymb}
\usepackage{graphicx}
\usepackage{dcolumn}
\usepackage{bm}
\usepackage{color}

\begin{document}
%
\title{Diffusion-mediated geminate reactions under excluded volume interactions} 
\author{Kazuhiko Seki}
\affiliation{
National Institute of Advanced Industrial Science and Technology (AIST)\\
AIST Tsukuba Central 5, Higashi 1-1-1, Tsukuba, Ibaraki, Japan, 305-8565
}
\author{Mariusz Wojcik}
\affiliation{
Institute of Applied Radiation Chemistry, Technical University of Lodz,
Wroblewskiego 15, 93-590 Lodz, Poland
}
\author{M. Tachiya
}
%
\affiliation{
National Institute of Advanced Industrial Science and Technology (AIST)\\
AIST Tsukuba Central 5, Higashi 1-1-1, Tsukuba, Ibaraki, Japan, 305-8565
}

\begin{abstract}

In this paper, influence of crowding by inert particles on 
the geminate reaction kinetics is theoretically investigated.     
Time evolution equations for the survival probability of a geminate pair 
are derived from the master equation taking into account the correlation among all diffusing particles 
and 
the results are compared with those obtained by Monte-Carlo simulations. 
In general, excluded volume interactions by the inert particles 
slow down the diffusive motion of reactants. 
However, 
when the initial concentration of the inert particles is uniform and high, 
we show that 
additional influence of interference between reaction and correlated diffusion 
accelerates the transient decay of the survival probability in the diffusion controlled limit. 
We also study the escape probability 
for a non-uniform initial distribution of the inert particles 
by taking the continuous limit in space. 
We show 
that reaction yield is increased when the reaction proceeds in the presence of 
a positive density gradient of the inert particles which inhibits the escape of reactants. 
The effect can be interpreted as a cage effect. 
\end{abstract}
\maketitle
\newpage
\setcounter{equation}{0}
\section{Introduction}
\vspace{0.5cm}

The importance of molecular crowding on chemical reactions has attracted great attention 
 in connection with biochemical reactions in living cells. \cite{Ellis,Schnell,Minton,Zimmerman,Zhou,Kapral,Yethiraj}
Living cells contain a high volume fraction of macromolecules, in addition to reactants. 
Although these macromolecules are not reactive, 
the excluded volume interactions between reactants and macromolecules significantly affect transport properties of reactants, and therefore biochemical reactions. 

In this paper, 
 we consider a fundamental reaction process called geminate reaction, which is observed in many systems 
 including those encountered in biology.  
In geminate reaction,  a pair of reactants is generated simultaneously 
and subsequently diffuse and react when they encounter.   
Geminate reactions are influenced by spatial diffusion of a pair and 
the intrinsic recombination rates. 
The influence of many body interactions between inert species and reactants on  
the geminate reaction kinetics     
can be very complicated and difficult to treat theoretically. 
The simplest model could be to assume 
that the reactants and the inert particles have the same size. 
Even under such simplification, 
the many body nature of the problem remains since 
the migration of reactive species correlates with the time dependent positions of inert species;   
the problem is still difficult to solve analytically. 

In order to retain many body nature in the simplest situation, 
we study geminate reaction between a static species and a diffusive species on a lattice. 
Reaction takes place according to the distance 
between one of the pair of reactants at the origin and the other. 
Inert particles perform random walks on a lattice. 
The transition to neighboring lattice sites is constrained by 
prohibiting the double occupancy; 
each lattice site can be occupied at most by a single diffusive particle regardless of 
whether it is reactive or inert. 
Particles are assumed to move randomly on vacancy sites of a lattice. 
However, 
the problem is still hard to solve analytically without approximations. 
Therefore, we perform Monte-Carlo simulations to evaluate approximations and 
to elucidate effects which cannot be studied analytically. 
In order to facilitate comparisons between theoretical results and those of simulations, 
the problem is further simplified;  
the origin is also allowed to be occupied at most by a single diffusive particle regardless of 
whether it is reactive or inert and 
reaction takes place according to the intrinsic reaction rate 
when the origin is occupied by a reactant. 

The excluded volume interactions were theoretically treated by 
Nakazato and Kitahara in tracer diffusion on a lattice. \cite{Nakazato}
Nakazato-Kitahara's formula of tracer diffusion constant interpolates between low and high concentrations of host particles and its accuracy 
is confirmed by comparison to the results of numerical simulations. \cite{Nakazato,Suzuki,Okamoto07,vanBeijeren85}

For target reactions where a static reactive particle (target) is surrounded by many reactive counterparts (quenchers), 
Nakazato-Kitahara's theory was successfully applied to calculate  
the survival probability of a target with a constraint of prohibited double occupancy of diffusing reactants. \cite{SekiPRE,SekiJCP}
It turned out that  
the decay of the target survival probability is accelerated by prohibiting the double occupancy. \cite{SekiPRE,SekiJCP}
Similar acceleration of the decay was obtained by  
other numerical and analytical approaches. 
\cite{Bhatia,Arora,Burlatsky,Sokolov,Lee2000,Shin2003,Zumofen_EV,Zhou91,Dong,Dzubiella,Sun,Dorsaz}
The acceleration of the decay is understood by noticing that 
the number of sites occupied by mobile reactants is generally larger at any time 
under the constraint of prohibited double occupancy at each lattice site.  \cite{SekiPRE,SekiJCP}
Accordingly, 
the probability of reaction between the target and a quencher 
is higher at any time when multiple occupancy is not allowed. 

Contrary to the target reaction, 
only a pair of reactants should be considered for geminate reaction. 
In other words, the number of sites occupied by reactants is not affected by prohibiting 
the double occupancy. 
However, the site blocking effects among diffusing particles influence the kinetics of geminate reaction 
through different mechanism from that in the case of target reactions. 
First, crowding of inert particles slows down diffusion of reactive particles and retards the reaction between the pair. 
Indeed, numerical simulations 
show that the reaction between a pair proceeds slowly by
the crowding of inert particles. \cite{Schmit}

In this paper, 
we 
study more comprehensively  the reaction of a pair of reactants 
performing diffusion under the constraint of prohibited double occupancy in the presence of inert particles. 
On the basis of results derived, the mechanism of site blocking effects by inert particles 
in the geminate reaction   
is investigated in detail when the initial distribution of inert particles is uniform. 
We show that reactions are influenced through excluded volume interactions 
not only through slowing down of diffusion of reactants but an interference between reaction and correlated diffusion  
in the diffusion-controlled limit. 
We also study the influence of inhomogeneous initial distributions of inert particles 
by taking the continuum limit in space. 
We show that the overall reaction yield is increased (decreased) from 
that assuming homogeneous initial distribution of inert particles by a positive (negative) 
density gradient of the inert particles. 

In Sec. II we formulate the problem in the case when the initial distribution of inert particles is uniform and 
derive the solution within the mean field approximation. 
In Sec. III higher order corrections to the mean field results are presented. 
The details of calculation are given in Appendix \ref{appA}. 
In Sec. IV, we compare the analytical results to simulation results. 
In Sec. V, the influence of non-uniform initial distribution of inert particles 
is investigated.  
The escape probability is derived for this case by using 
the continuous limit derived in Appendix \ref{appB}. 
Section VI is devoted for conclusions. 

\section{Geminate pair reaction under the presence of inert particles}
\vspace{0.5cm}
For simplicity, 
we formulate the problem on a lattice where a reactive particle and inert particles perform random walks. 
The particles are assumed to move randomly on the vacant sites of a lattice. 
One of the reactants of the pair does not move and 
its position is taken as 
the origin of the coordinate system. 
The reactive particle undergoes reaction according to the distance from the origin, $r$. 
We denote the intrinsic reaction rate by $k(r)$. 

The tracer-diffusion in concentrated lattices was studied by Nakazato and Kitahara 
in the absence of reaction. \cite{Nakazato} 
The diffusion of the tagged particle in the presence of site blocking by other particles 
has been studied. \cite{Nakazato,Suzuki,Okamoto07,vanBeijeren85}
Following them, we  introduce ket vectors to show occupancy of a site by diffusing particles. 
The ket vector $| \vec{r}, \bullet \rangle$ denotes the occupation 
of site $\vec{r}$ 
by a reactive particle, 
the ket vector $| \vec{r}, \circ \rangle$ denotes the occupation 
of site $\vec{r}$ 
by a inert particle, 
and $| \vec{r}, \phi \rangle$ represents that site $\vec{r}$ is empty. 
The conditional probability of finding inert particles at $(\vec{f}_1, \vec{f}_2, \cdots, \vec{f}_N)$ 
and the reactant at $\vec{n}$ at time $t$ when 
the initial configuration of inert particles is $(\vec{i}_1, \vec{i}_2, \cdots, \vec{i}_N)$ and that of the reactant is $\vec{m}$ 
is written as, 
\begin{align}
P(\vec{f}_1, \vec{f}_2, \cdots , \vec{f}_N,\vec{n}, t ; \vec{i}_1, \vec{i}_2, \cdots , \vec{i}_N,\vec{m}) 
=& 
\left( \prod_{\ell=1}^N \langle \vec{f}_\ell, \circ | \right) \left( \prod_{\ell=N+1}^M \langle \vec{f}_\ell, \phi | \right) \langle \vec{n}, \bullet |
\exp({\cal L} t) 
\nonumber \\
&| \vec{m}, \bullet \rangle \left( \prod_{\ell=1}^N | \vec{i}_\ell, \circ \rangle \right) \left( \prod_{\ell=N+1}^M | \vec{i}_\ell, \phi \rangle \right) , 
\label{cond_prob}
\end{align} 
where $N$ and $M$ denote the numbers of inert particles and lattice sites, respectively. 
${\cal L}$ is given by the sum of the term describing diffusion 
$L_{\rm w}$ and that describing reaction $L_{\rm rc}$, 
${\cal L}=L_{\rm w} + L_{\rm rc}$.  
$L_{\rm w}$ is explicitly expressed as, \cite{Nakazato,Suzuki,Okamoto07}
\begin{align}
L_{\rm w} =& \gamma_B \sum_{\langle n,m \rangle}
\left( 
|\vec{r}_n, \bullet \rangle \langle \vec{r}_n, \phi | 
\cdot 
|\vec{r}_m, \phi \rangle \langle \vec{r}_m, \bullet | 
- |\vec{r}_n, \bullet \rangle \langle \vec{r}_n, \bullet | 
\cdot 
|\vec{r}_m, \phi \rangle \langle \vec{r}_m, \phi | 
\right) + \nonumber \\
& \gamma_w \sum_{\langle n,m \rangle}
\left( 
|\vec{r}_n, \circ \rangle \langle \vec{r}_n, \phi | 
\cdot 
|\vec{r}_m, \phi \rangle \langle \vec{r}_m, \circ | 
- |\vec{r}_n, \circ \rangle \langle \vec{r}_n, \circ | 
\cdot 
|\vec{r}_m, \phi \rangle \langle \vec{r}_m, \phi | 
\right) ,
\end{align}
where the sum is taken over all nearest neighbor pairs of 
the accessible lattice sites by   
the diffusing particles. 
$\gamma_B$ is given by 
$\gamma_B=\Gamma_B/(2d)$, where $\Gamma_B$ is the jump frequency of a reactive particle and 
$d$ denotes the lattice dimension. 
Similarly, we define $\gamma_w=\Gamma_w/(2d)$ where 
$\Gamma_w$ is the jump frequency of inert particles. 
$L_{\rm rc}$ describes the reaction from an occupied site $\vec{r}_n$ 
with the rate $k \left(\vec{r}_n \right)$, 
\cite{Doi,Kotomin,Peliti}
\begin{eqnarray}
L_{\rm rc} = - \sum_{n=1}^M k \left(\vec{r}_n \right) | \vec{r}_n, \bullet \rangle \langle \vec{r}_n, \bullet | .
\end{eqnarray}
The conditional probability,  
$P_N \left(\vec{n},t | \vec{m}, 0 \right)$,  
that the reactant is at site $\vec{n}$ at time $t$ when 
it was initially at $\vec{m}$ under the assumption of random initial occupation of inert particles is 
obtained from Eq. (\ref{cond_prob}) by multiplying $1/\left(_{M} C_N \right)$ and 
summing over all possible initial and final configurations of the inert particles. 
By defining the characteristic function by, 
\begin{align}
\phi(\vec{n},t | \vec{m}, 0; x) \equiv \sum_{N=0}^{M} P_N \left(\vec{n},t | \vec{m}, 0 \right) x^N , 
\label{def_charf}
\end{align}
it can be expressed as, 
\begin{align}
\phi(\vec{n},t | \vec{m}, 0; x) = \frac{\left(1+x\right)^M}{_MC_N} g \left(\vec{n},t | \vec{m}, 0 ; x \right)  , 
\label{charf1}
\end{align}
where we define, 
\begin{align}
g \left(\vec{n},t | \vec{m}, 0 ; x \right) \equiv \langle \left\{ \phi \right\} | \langle \vec{n}, \bullet|  \exp \left(\tilde{L} (\theta) t  \right) 
|\left\{ \phi \right\} \rangle  | \vec{m}, \bullet \rangle , 
\label{def_g}
\end{align}
$\tilde{L}(\theta)= \exp \left( - \theta S \right) {\cal L}  \exp \left( \theta S \right)$, 
$S\equiv \sum_{\ell=1}^M \left(  |\vec{r}_\ell, \circ \rangle \langle  \vec{r}_\ell, \phi |
- |\vec{r}_\ell, \phi \rangle  \langle  \vec{r}_\ell, \circ | 
\right) 
$, and 
$x=\tan^2 \theta$. 
It is convenient to introduce abbreviations,  
\begin{align}
\langle \left\{ \phi \right\} | \langle \vec{r}_j, \bullet| 
&\equiv \left( \prod_{\ell=1}^{M'} \langle \vec{r}_\ell, \phi |  \right) \langle \vec{r}_j, \bullet|  ,\\
|\left\{ \phi \right\} \rangle  | \vec{r}_j, \bullet \rangle 
&\equiv \left( \prod_{\ell=1}^{M'} | \vec{r}_\ell, \phi \rangle  \right) | \vec{r}_j, \bullet \rangle  ,
\end{align}
where $M'$ denotes that the site $\vec{r}_j$ is excluded in the product.

The inverse transformation is given  
by applying the Cauchy's integral theorem, 
\begin{eqnarray}
P_N \left(\vec{n},t | \vec{m}, 0 \right) = \frac{1}{2 \pi i} \int d\, x \frac{1}{x^{N+1}} \phi(\vec{n},t | \vec{m}, 0; x)  ,  
\label{def_invt}
\end{eqnarray} 
where the path of integration encircles the origin on the complex plane.

In the thermodynamic limit of $M \rightarrow \infty$ with $c=N/M$ being constant, 
the right hand side of Eq. (\ref{def_invt}) can be calculated by applying the saddle point method, 
\cite{Nakazato,Suzuki,Okamoto07}
\begin{align}
P \left(\vec{n},t | \vec{m}, 0 \right) = g \left(\vec{n},t | \vec{m}, 0 ; c/(1-c) \right) . 
\label{tdl}
\end{align}
The survival probability of a pair at time $t$ whose initial separation is given by $\vec{m}$ is defined by, 
\begin{align}
S \left(\vec{m}, t \right) = \sum_{\vec{n}} P \left(\vec{n},t | \vec{m}, 0 \right) . 
\label{def_suv}
\end{align}
From Eqs. (\ref{def_g}) and (\ref{tdl}) the Laplace transform of the survival probability, 
$\hat{S} \left(\vec{m}, s \right) = \int_0^\infty dt \exp (-st) S \left(\vec{m}, t \right)$, 
is expressed as, 
\begin{align}
\hat{S} \left(\vec{m}, s \right) = \sum_{\vec{n}} 
\langle \left\{ \phi \right\} | \langle \vec{n}, \bullet|  
\frac{1}{s- \tilde{L}(\theta^*)}
|\left\{ \phi \right\} \rangle  | \vec{m}, \bullet \rangle 
,  
\label{suv1}
\end{align}
where $\tan \theta^* = \sqrt{c/(1-c)}$. 
$\tilde{L}(\theta^*)$ can be expressed by the sum of the term describing diffusion and that describing reaction, 
$\tilde{L}(\theta^*)=\tilde{L}_{\rm rw}+\tilde{L}_{\rm rc}$. 
Even after the transformation the term describing reaction is not changed, 
$\tilde{L}_{\rm rc}=L_{\rm rc}$, 
while 
$\tilde{L}_{\rm rw}$ is given by the sum, 
$\tilde{L}_{\rm rw}= \tilde{L}_{\rm rw0} + \tilde{L}_{\rm rw1}$, where 
$\tilde{L}_{\rm rw0}$ describes the transition under the conservation constraint of the number of particles, 
\cite{Nakazato,Suzuki,Okamoto07}
\begin{align}
\tilde{L}_{\rm rw0} =& 
(1-c) \gamma_B \sum_{\langle n,m \rangle} 
\left( 
|\vec{r}_n, \bullet \rangle \langle \vec{r}_n, \phi | 
\cdot 
|\vec{r}_m, \phi \rangle \langle \vec{r}_m, \bullet | 
- |\vec{r}_n, \bullet \rangle \langle \vec{r}_n, \bullet | 
\cdot 
|\vec{r}_m, \phi \rangle \langle \vec{r}_m, \phi | 
\right) + \nonumber \\
& 
c \gamma_B  \sum_{\langle n,m \rangle} 
\left( 
|\vec{r}_n, \bullet \rangle \langle \vec{r}_n, \circ | 
\cdot 
|\vec{r}_m, \circ \rangle \langle \vec{r}_m, \bullet | 
- |\vec{r}_n, \bullet \rangle \langle \vec{r}_n, \bullet | 
\cdot 
|\vec{r}_m, \circ \rangle \langle \vec{r}_m, \circ | 
\right) + \nonumber \\
&
\gamma_w  \sum_{\langle n,m \rangle} 
\left( 
|\vec{r}_n, \circ \rangle \langle \vec{r}_n, \phi | 
\cdot 
|\vec{r}_m, \phi \rangle \langle \vec{r}_m, \circ | 
- |\vec{r}_n, \circ \rangle \langle \vec{r}_n, \circ | 
\cdot 
|\vec{r}_m, \phi \rangle \langle \vec{r}_m, \phi | 
\right) , 
\label{def_L0}
\end{align}
and $\tilde{L}_{\rm rw1}$ describes the transition where the number of particles is not conserved, 
\cite{Nakazato,Suzuki,Okamoto07}
\begin{align}
\tilde{L}_{\rm rw1}  =&\sqrt{c(1-c)}  \, \gamma_B  \sum_{\langle n,m \rangle} 
\left( 
|\vec{r}_n, \bullet \rangle \langle \vec{r}_n, \bullet | 
\cdot 
|\vec{r}_m, \phi \rangle \langle \vec{r}_m, \circ | 
+ \right. \nonumber \\
&
\vec{r}_n, \bullet \rangle \langle \vec{r}_n, \bullet | 
\cdot 
|\vec{r}_m, \circ \rangle \langle \vec{r}_m, \phi | 
- 
|\vec{r}_n, \bullet \rangle \langle \vec{r}_n, \phi | 
\cdot 
|\vec{r}_m, \circ \rangle \langle \vec{r}_m, \bullet | 
- \nonumber \\
&
\left.
|\vec{r}_n, \bullet \rangle \langle \vec{r}_n, \circ | 
\cdot 
|\vec{r}_m, \phi \rangle \langle \vec{r}_m, \bullet | 
\right) .
\label{def_L1}
\end{align}

By introducing the identity, 
\begin{align}
\frac{1}{s-\tilde{L}(\theta^*)}
= \frac{1}{s} \left( 
1+ \frac{1}{s-\tilde{L}(\theta^*)} \tilde{L}(\theta^*)
\right),  
\end{align}
Eq, (\ref{suv1}) can be rewritten as, 
\begin{align}
s \hat{S} \left(\vec{m}, s \right) = 1+ \sum_{\vec{n}} 
\langle \left\{ \phi \right\} | \langle \vec{n}, \bullet|  
\frac{1}{s- \tilde{L}(\theta^*)}\tilde{L}(\theta^*) 
|\left\{ \phi \right\} \rangle  | \vec{m}, \bullet \rangle 
. 
\label{suv2}
\end{align}

In the lowest order approximation, 
the perturbation term, $\tilde{L}_{\rm rw1}$, is ignored in the numerator of
Eq. (\ref{suv2}) and we obtain, 
\begin{align}
s \hat{S} \left(\vec{m}, s \right) = 1+ \sum_{\vec{n}} 
\langle \left\{ \phi \right\} | \langle \vec{n}, \bullet|  
\frac{1}{s- \tilde{L}(\theta^*)}\left( \tilde{L}_{\rm rw0} +  \tilde{L}_{\rm rc} \right)
|\left\{ \phi \right\} \rangle  | \vec{m}, \bullet \rangle 
. 
\label{suv3}
\end{align}
By using Eq. (\ref{suv1})  and the fact that the number of particles is conserved for 
both  $\tilde{L}_{\rm rw0}$ and $\tilde{L}_{\rm rc}$, Eq. (\ref{suv3}) leads to 
\begin{align}
s \hat{S} \left(\vec{m}, s \right)-1  = \gamma_B (1-c) \sum_{j} 
\left( \hat{S} \left(\vec{m}+\vec{b}_j, s \right) -\hat{S} \left(\vec{m}, s \right)
\right)
- k \left( \vec{m} \right) \hat{S} \left(\vec{m}, s \right) 
, 
\label{suv4}
\end{align}
where $\vec{m}+\vec{b}_j$ denotes a nearest neighbor of the site $\vec{m}$ 
and the sum is taken over all nearest neighbor sites. 
By the inverse Laplace transformation, the equation for the survival probability at time $t$ 
of a pair with initial separation $\vec{m}$  is 
obtained, 
\begin{align}
\frac{\partial}{\partial t} S \left(\vec{m}, t \right)  = \gamma_B (1-c) \sum_{j} 
\left( S \left(\vec{m}+\vec{b}_j, t \right) -S \left(\vec{m}, t \right)
\right)
- k \left( \vec{m} \right) S \left(\vec{m}, t \right) 
. 
\label{suv5}
\end{align}
In the lowest order approximation, 
the site blocking effects by inert particles reduces the transition rate. 
The transition rate is reduced since  jump to a neighboring site is allowed only when 
the neighboring site is empty. 
The vacant probability is $1-c$  in the mean field picture. 
Equation (\ref{suv5}) is a mean-field result in the sense that
the reduction factor is given by $1-c$. 
The transition rate of the reactant particle decreases linearly with increasing the concentration of inert particles.  

For localized reactions, $k \left( \vec{m} \right) = k_0 \delta_{\vec{m},\vec{0}}$, 
the general solution after the Laplace transformation is obtained as, 
\begin{align}
\hat{S} \left(\vec{m}, s \right) = \frac{1}{s} \left( 1 - \frac{\hat{G}_0 (\vec{m},s) k_0 }{1+ 
\hat{G}_0 (\vec{0},s) k_0 } \right) , 
\label{suv_invFourier_sol}
\end{align}
where 
the Green's function, 
\begin{align}
\hat{G}_0 (\vec{j}, s) &= \frac{1 - \hat{\psi}_B (s)}{s} U(\vec{j}, s) ,
\label{G0}
\end{align}
 is given in terms of 
the lattice Green's function, \cite{Hughes}
\begin{align}
U(\vec{j}, s) &= \frac{1}{(2\pi)^d} \int \cdots \int_{-\pi}^{\pi} d^d \vec{k}
\frac{\exp \left( 
-i \vec{k} \cdot \vec{j} 
\right)}{1 - \hat{\psi}_B (s) \lambda (\vec{k})}, 
\label{latticeU}
\end{align}
where $\hat{\psi}_B (s)$ is given by $\hat{\psi}_B (s)=\Gamma_B(1-c)/(s+ \Gamma_B (1-c))$, 
the structure factor is defined by 
$\lambda (\vec{k}) \equiv \frac{1}{2d} \sum_{j=1}^{2d} \cos \left(\vec{k} \cdot \vec{b}_j/b \right)$ and  
$b$ denotes the lattice spacing.

The recombination probability of a particle starting from $\vec{m}$, $\kappa \left( \vec{m} \right)=1-\lim_{t\rightarrow \infty} S \left(\vec{m}, t \right) $,  
is obtained as
\begin{align}
\kappa \left( \vec{m} \right) = \frac{\displaystyle U(\vec{m}, 0)}
{\displaystyle \frac{\Gamma_B(1-c)}{k_0} +  U(\vec{0}, 0)}  .
\label{RPl}
\end{align}
Note that $U(\vec{m}, 0)$ for any $\vec{m}$ is independent of the concentration of the inert particles, $c$,  
in the mean-field result. 
In the limit of perfectly absorbing boundary condition ($k_0 \rightarrow \infty$), 
the recombination probability is independent of the concentration of inert particles.
For partially absorbing boundary conditions 
the recombination probability increases by increasing the concentration of the inert particles. 
The escape probability, $\lim_{t\rightarrow \infty} S \left(\vec{m}, t \right) $, 
which is defined as the probability of a pair with initial separation $\vec{m}$ surviving at infinite time, is given by, 
\begin{align}
\varphi \left( \vec{m} \right) = 1 - \kappa \left( \vec{m} \right) . 
\label{esc}
\end{align}

We have derived the simplest results on the survival probability of a geminate pair 
by ignoring correlations higher than the two-point correlation between the initial position and 
the position at an arbitrary time. 
In the reaction-diffusion equation thus derived, 
the presence of the inert particles only reduces the diffusion coefficient 
linearly with increasing the concentration of the inert particles and the diffusion and the reaction do not interfere. 
In the subsequent section, we show that 
the diffusion and the reaction interfere in the presence of inert particles 
if we consider higher order correlations.

\section{Correction to the mean field equation}
\vspace{0.5cm}
If we ignore correlations higher than two-point correlations, 
the Bardeen-Herring back correlation is not taken into account. \cite{Bardeen} 
The Bardeen-Herring back correlation takes place when 
the reactant 
hops to a vacant site, 
leaving the previous occupied site vacant; 
after the hopping 
the transition probability of the reactant 
back to the previously occupied site is higher than other sites.  
The velocity autocorrelation function in a lattice gas 
shows a long time tail with a negative value due to
the Bardeen-Herring back correlation.  \cite{vanBeijeren85}
Suppose that a reactant occupies a reactive site after a hopping. 
The rate of hopping back to the previous site competes with that of reaction. 
In this way, the reaction interferes the correlated diffusion. 
Interference means that the reaction process and the diffusion are not statistically independent. 
In this section, we study the interference between the reaction and the correlated diffusion 
by taking into account 
the Bardeen-Herring back correlation. 
As in the previous section, 
we assume the initial uniform distribution for the inert particles. 

The effect of the interference between the reaction and the correlated diffusions 
can be calculated as the correction to 
the simple diffusion-reaction equation, 
Eq. (\ref{suv3}). 
The exact relation, Eq. (\ref{suv2}), can be rewritten as, 
\begin{align}
s \hat{S} \left(\vec{m}, s \right) = 1+ \sum_{\vec{n}} 
\langle \left\{ \phi \right\} | \langle \vec{n}, \bullet|  
\frac{1}{s- \tilde{L}(\theta^*)}\left( \tilde{L}_{\rm rw0} +  \tilde{L}_{\rm rc} \right)
|\left\{ \phi \right\} \rangle  | \vec{m}, \bullet \rangle 
+\hat{R}(\vec{m},s) 
, 
\label{suve1}
\end{align}
where $\hat{R}(\vec{m},s)$ represents the correction to Eq. (\ref{suv3}) and is given by, 
\begin{align}
\hat{R}(\vec{m},s) =\sum_{\vec{n}} \langle \left\{ \phi \right\} | \langle \vec{n}, \bullet|  
\frac{1}{s- \tilde{L}(\theta^*)}\tilde{L}_{\rm rw1} 
|\left\{ \phi \right\} \rangle  | \vec{m}, \bullet \rangle . 
\label{suve2}
\end{align}
By noticing 
$\tilde{L}_{\rm rw1} =\tilde{L}(\theta^*)-\tilde{L}_{\rm rw0} -  \tilde{L}_{\rm rc}$, 
we can prove the operator identity, 
\begin{align}
\frac{1}{s- \tilde{L}(\theta^*)} =
\frac{1}{s- \tilde{L}(\theta^*)}\tilde{L}_{\rm rw1} 
\frac{1}{s- \tilde{L}_{\rm rw0} -  \tilde{L}_{\rm rc}} 
+
\frac{1}{s- \tilde{L}_{\rm rw0} -  \tilde{L}_{\rm rc}}.  
\label{suve3}
\end{align}
$\tilde{L}_{\rm rw0}$ and $\tilde{L}_{\rm rc}$ conserve the number of $\bullet$ 
in the bra and ket notations, 
while $\tilde{L}_{\rm rw1}$ does not, 
so we have 
\begin{align}
\langle \left\{ \phi \right\} | \langle \vec{n}, \bullet| 
\frac{1}{s- \tilde{L}_{\rm rw0}-  \tilde{L}_{\rm rc}}\tilde{L}_{\rm rw1} 
|\left\{ \phi \right\} \rangle  | \vec{m}, \bullet \rangle =0 . 
\label{suve4}
\end{align}
If we substitute Eq. (\ref{suve3}) into Eq. (\ref{suve2}) and   
use Eq. (\ref{suve4}),  
Eq. (\ref{suve2}) can be expressed as,   
\begin{align}
\hat{R}(\vec{m},s) &=\sum_{\vec{n}} \langle \left\{ \phi \right\} | \langle \vec{n}, \bullet|  
\frac{1}{s- \tilde{L}(\theta^*)}\tilde{L}_{\rm rw1} 
\frac{1}{s- \tilde{L}_{\rm rw0}-  \tilde{L}_{\rm rc}}\tilde{L}_{\rm rw1} 
|\left\{ \phi \right\} \rangle  | \vec{m}, \bullet \rangle \\
&= \sum_{\vec{n}} \hat{S} \left(\vec{n}, s \right) 
 \langle \left\{ \phi \right\} | \langle \vec{n}, \bullet| 
 \tilde{L}_{\rm rw1} 
\frac{1}{s- \tilde{L}_{\rm rw0}-  \tilde{L}_{\rm rc}}\tilde{L}_{\rm rw1} 
|\left\{ \phi \right\} \rangle  | \vec{m}, \bullet \rangle
 ,  
\label{suve5}
\end{align}
where the definition of $\hat{S} \left(\vec{n}, s \right)$ given by 
Eq. (\ref{suv1}) is substituted. 
By introducing the explicit expression of $\tilde{L}_{\rm rw1}$ given by Eq. (\ref{def_L1}), 
we obtain, 
\begin{multline}
\langle \left\{ \phi \right\} | \langle \vec{n}, \bullet| 
 \tilde{L}_{\rm rw1} 
\frac{1}{s- \tilde{L}_{\rm rw0}-  \tilde{L}_{\rm rc}}\tilde{L}_{\rm rw1} 
|\left\{ \phi \right\} \rangle  | \vec{m}, \bullet \rangle 
=  \\
\gamma_B^2 c (1-c) \sum_r \sum_q 
 \left[
 G\left(\vec{n}\bullet ,\vec{n} + \vec{b}_r \circ |  \vec{m} \bullet, \vec{m}+\vec{b}_q \circ  ,s \right)
 -G\left(\vec{n}+ \vec{b}_r \bullet ,\vec{n}  \circ | \vec{m} \bullet,  \vec{m}+\vec{b}_q \circ  ,s \right) 
\right.  \\ \left.
 -G\left(\vec{n} \bullet ,\vec{n} + \vec{b}_r \circ |   \vec{m}+\vec{b}_q \bullet, \vec{m} \circ ,s \right)
 +G\left(\vec{n}+ \vec{b}_r \bullet ,\vec{n}  \circ |  \vec{m}+\vec{b}_q \bullet, \vec{m} \circ  ,s \right)
 \right] .  
 \label{suve6}
\end{multline}
Here we define the four-point correlation function as, 
\begin{align}
 G\left(\vec{r}_i \bullet ,\vec{r}_j \circ |  \vec{r}_k \bullet, \vec{r}_\ell \circ  ,s \right)
 = \langle \left\{ \phi \right\} | \langle \vec{r}_i, \bullet| 
\langle \vec{r}_j, \circ| 
\frac{1}{s- \tilde{L}_{\rm rw0}-  \tilde{L}_{\rm rc}}
|\left\{ \phi \right\} \rangle  | \vec{r}_k, \bullet \rangle  | \vec{r}_\ell, \circ \rangle
\label{suve7}
\end{align}
using the abbreviation,  
\begin{align}
\langle \left\{ \phi \right\} | \langle \vec{r}_i, \bullet| 
\langle \vec{r}_j, \circ| 
&\equiv \left( \prod_{\ell=1}^{M''} \langle \vec{r}_\ell, \phi |  \right) \langle \vec{r}_i, \bullet| \langle \vec{r}_j, \circ| ,
\label{suve8_1}\\
|\left\{ \phi \right\} \rangle  | \vec{r}_i, \bullet \rangle  | \vec{r}_j, \circ \rangle
&\equiv \left( \prod_{\ell=1}^{M''} | \vec{r}_\ell, \phi \rangle  \right) | \vec{r}_i, \bullet \rangle | \vec{r}_j, \circ \rangle ,
\label{suve8}
\end{align}
where $M''$ denotes that the site $\vec{r}_i$ and the site $\vec{r}_j$ are excluded in the product. 
Eqs. (\ref{suve8_1})-(\ref{suve8}) represent the state 
that all sites are vacant 
except the site $\vec{r}_i$ occupied by the reactant and 
the site $\vec{r}_j$ occupied by an inert particle. 
By substituting Eqs. (\ref{suve5}) and (\ref{suve6}) in Eq. (\ref{suve1}), 
we have, 
\begin{align}
s \hat{S} \left(\vec{m}, s \right)-1  = \gamma_B (1-c) \sum_{\vec{n}} \sum_{r} 
F_c(\vec{n},\vec{m},\vec{b}_r,s) \left( \hat{S} \left(\vec{n}+\vec{b}_r, s \right) -
\hat{S} \left(\vec{n}, s \right)
\right) 
- k \left( \vec{m} \right) \hat{S} \left(\vec{m}, s \right) 
, 
\label{suve9}
\end{align}
where the kernel $\hat{F}_c(\vec{n},\vec{m},\vec{b}_r,s)$ is given by 
\begin{multline}
\hat{F}_c(\vec{n},\vec{m},\vec{b}_r,s) = \delta_{\vec{n},\vec{m}} - \gamma_B c  \sum_q 
 \left[G\left(\vec{n} \bullet ,\vec{n}+\vec{b}_r \circ |  \vec{m} \bullet,  \vec{m}+\vec{b}_q \circ  ,s \right)
 \right. \\
 \left. 
  -G\left(\vec{n} \bullet ,\vec{n}+\vec{b}_r \circ |  \vec{m}+\vec{b}_q \bullet,  \vec{m} \circ  ,s \right)
 \right] . 
\label{suve10}
\end{multline}
By the inverse Laplace transformation of Eq. (\ref{suve9}), the survival probability 
is shown to satisfy the diffusion-reaction equation 
in which the diffusion term is expressed by the time convolution with the nonlocal kernel, 
$F_c(\vec{n},\vec{m},\vec{b}_r,t)$. 

The time convolution represents the memory effect originating from the correlation 
between the mobile reactant and the inert particles. 
Reactant motion is correlated with the time dependent arrangements of the inert particles 
through prohibited double occupancy of the lattice sites. 
In particular, 
the site occupied by the reactant becomes empty just after the hopping of the reactant and 
the chance of back transition to the previously occupied site is high. 
The back transition probability of the reactant decreases as time proceeds 
because the empty site generated by the hop of the reactant to a vacant site may 
be occupied by another inert particle. 
The time dependence of  back-jump correlation is the origin of the memory effect. 

In principle, 
the back-jump correlation competes with reaction. 
Suppose that the reactant hops to the reactive site. 
The probability of jump back to the previously occupied site decreases as the reaction rate increases. 
Since 
Eq. (\ref{suve7}) includes the operator describing reaction, $\tilde{L}_{\rm rc}$, 
the diffusion memory kernel,
$\hat{F}_c(\vec{n},\vec{m},\vec{b}_r,s)$, depends on the reaction rate. 
In order to obtain $\hat{F}_c(\vec{n},\vec{m},\vec{b}_r,s)$  
we need to solve an equation for  $
G\left(\vec{n} \bullet ,\vec{n}+\vec{b}_r \circ |  \vec{m} \bullet, \vec{r} \circ  ,s \right)$.

In Eq. (\ref{suve12}) of Appendix \ref{appA}, 
we show that the equation for  $G\left(\vec{n} \bullet ,\vec{n}+\vec{b}_r \circ |  \vec{m} \bullet, \vec{r} \circ  ,s \right)$ includes the reactive sink term. 
The interference between reaction and correlated diffusion is taken into account 
by the four-point correlation function. 
In the simplest theory given by the two-point function, Eq. (\ref{suv5}),  
the interference between reaction and correlated diffusion is not taken into account. 

When the reactive sink strength changes according to the distance from the origin, 
the diffusion term given in terms of the four-point correlation function 
depends on the distance from the origin accordingly. 
In addition, the presence of the inert particles gives rise to correlation over distances as a result of the excluded volume interactions 
between the inert particles and the reactant. 
Interference between reaction and the correlated diffusion beaks down the translational invariance as shown in 
Eq. (\ref{suve12}) of Appendix \ref{appA} and 
the resultant equation is hard to solve. 
In the next section, we use numerical simulations to 
study the interference effect.

When we ignore the interference between reaction and 
correlated diffusion, 
the translation invariance is satisfied for the four-point correlation functions. 
Under the translational invariance, 
$ G\left(\vec{n} \bullet ,\vec{n}+\vec{b}_r \circ |  \vec{m} \bullet, \vec{r} \circ  ,s \right)$ depends only 
on relative vectors and satisfies, 
\begin{align}
G\left(\vec{n} \bullet ,\vec{n}+\vec{b}_r \circ |  \vec{m} \bullet, \vec{r} \circ  ,s \right)
=  G\left(\vec{n}\,' \bullet ,\vec{n}\,'+\vec{b}_r \circ |  \vec{0} \bullet, \vec{r}\,' \circ  ,s \right),  
\label{suve13p}
\end{align}
where $\vec{n}\,'=\vec{n}-\vec{m}$ and $\vec{r}\,'=\vec{r}-\vec{m}$. 
Since the number of independent variables is reduced, 
it is convenient to introduce a new notation, 
\begin{align}
G^{(T)} \left(\vec{n}\,' \bullet ,\vec{b}_r \circ | \vec{r}\,' \circ  ,s \right)
=
 G\left(\vec{n}\,' \bullet ,\vec{n}\,'+\vec{b}_r \circ |  \vec{0} \bullet, \vec{r}\,' \circ  ,s \right). 
\label{suve13}
\end{align}
In this case, the time evolution equation for the survival probability is expressed after the spatial Fourier transform as, 
\begin{align}
\frac{\partial}{\partial t} S \left(\vec{k}, t \right) = \int_0^t d t_1 M(k, t -t_1)  
S \left(k, t_1 \right) -
 \sum_{\vec{m}} \exp \left(i \vec{k} \cdot \vec{m} \right) k \left( \vec{m} \right) S \left(\vec{m}, t \right) 
. 
\label{suve14}
\end{align}
In the Laplace domain,  $\hat{M}(k, s)$ can be regarded as a self-energy or memory function and is expressed as, 
\begin{align}
\hat{M}(\vec{k}, s )&=-\Gamma_B (1-c)\left(1-\lambda (\vec{k})  \right) + \gamma_B^2 c (1-c)
\hat{T}_c \left(\vec{k},s\right)  ,
\label{suve15}
\end{align}
where the correlations among the inert particles and the reactant 
as a result of the excluded volume interactions are included in, 
\begin{align}
\hat{T}_c(\vec{k},s) &= \sum_q \sum_r 
\left(1 - \exp \left(-i \vec{k} \cdot \vec{b}_r \right) \right) \tilde{G} \left( \vec{k} ,  {\vec{b}_r}\,|\vec{b}_q ,s \right)
\left(1 - \exp \left(i \vec{k} \cdot \vec{b}_q \right) \right) , 
\label{suve16} \\
\tilde{G} \left(\vec{k} , {\vec{b}_r}\,| \vec{r} ,s \right)&= \sum_{\vec{\ell}} \exp \left( i \vec{k}\cdot \vec{\ell} \right) 
G^{(T)} \left(\vec{\ell}\, ,{\vec{b}_r}\, | \vec{r} ,s \right) .
\label{suve16_1}
\end{align}
The same form of memory function expressed in terms of 
the four-point correlation function was derived by a different method. \cite{Tahir-Kheli} 
The equation for $\tilde{G} \left(\vec{k} , {\vec{b}_r}\,|  \vec{r} ,s \right)$ is 
explicitly shown in Eq. (\ref{suve18_B5}) of Appendix \ref{appA}. 

In the limit of small wavelength, $k \rightarrow 0$, 
Eq. (\ref{suve14}) can be expressed after the inverse Laplace transformation as, 
\begin{align}
\frac{\partial}{\partial t} S \left(\vec{m}, t \right)  = D_B (1-c) \sum_{j} 
\int_0^t \, dt_1 f_c \left(t-t_1 \right) 
\nabla^2 
S \left(\vec{m}, t_1 \right)
- k \left( \vec{m} \right) S \left(\vec{m}, t \right) 
, 
\label{suve18}
\end{align}
where the diffusion constant is defined by, 
$D_B=b^2 \gamma_B$,  
the correlation factor in the Laplace domain is given 
for the hypercubic lattices by, 
\begin{align}
\hat{f}_c(s) = 1 - \gamma_B c \sum_r 
 \left[G^{(a)} \left( {\vec{b}_r}\,|  \vec{b}_r ,s \right)- G^{(a)} \left({\vec{b}_r}\,|   -\vec{b}_r ,s \right)
  \right] , 
\label{suve19}
\end{align}
and the initial condition is $S \left(\vec{m}, t=0 \right)=1$. 
$G^{(a)} \left( {\vec{b}_r}\,|  \vec{b}_q ,s \right)$ is defined by, 
\begin{align}
G^{(a)} \left({\vec{b}_r}\,|  \vec{b}_q ,s \right) 
 =\sum_{\vec{m}} \langle \left\{ \phi \right\} | \langle \vec{n}, \bullet| 
\langle \vec{n}+\vec{b}_r, \circ| 
\frac{1}{s- \tilde{L}_{\rm rw0}}
|\left\{ \phi \right\} \rangle  | \vec{m}, \bullet \rangle  | \vec{m}+\vec{b}_q, \circ \rangle ,  
\label{suve20}
\end{align}
which is independent of the choice of $\vec{n}$ as shown in 
Eq. (\ref{suve22}) of 
Appendix \ref{appA}. 
The equation for $G^{(a)} \left( {\vec{b}_r}\,|  \vec{b}_q ,s \right)$ is known and has been studied 
to obtain the tracer-diffusion coefficient. \cite{Nakazato,Suzuki,Okamoto07}
It is shown in Eq. (\ref{suve22}) of Appendix \ref{appA}. 
Its solution is known and the correlation factor can be expressed as, 
\begin{align}
\hat{f}_c(s) = \frac{1-\mu(s)}{\displaystyle 1 - \mu(s) 
\frac{\gamma_w+\gamma_B(1-3 c)}{\gamma_w+\gamma_B(1-c)}}, 
\label{suve23}
\end{align}
where $\mu(s)$ is given by, 
\begin{align}
\mu(s) = \frac{1}{(2\pi)^d} \int \cdots \int_{-\pi}^{\pi} d^d \vec{k}
\frac{2\sin^2 k_1}{ 
(s/\gamma_t)+ 2d[1- \lambda (\vec{k})]} , 
\label{suve24}
\end{align}
and $\gamma_t= \gamma_w+\gamma_B(1-c)$. 
In the original derivation of the correlation factor $\hat{f}_c(s)$ and the tracer-diffusion coefficient by Nakazato and Kitahara, 
a projection operator method is applied. 
Here, a reaction-diffusion equation for the survival probability is derived directly without using the projection operator method. 

In order to take into account the memory effect on the transient decay of the survival probability 
with reasonable simplicity, Eqs. (\ref{suv_invFourier_sol})-(\ref{latticeU}) are used with 
\begin{align}
\hat{\psi}_B (s) = \frac{\Gamma_B \hat{f}_c(s)}{s+ \Gamma_B \hat{f}_c(s) (1-c)} . 
\label{psiBs}
\end{align}
In this approximation, 
the lattice Green function valid for finite wave length is used together with the expression of 
$\hat{f}_c(s)$ derived in the limit of $k \rightarrow 0$. 
In the subsequent section, we show  by comparison with 
simulation results that the approximation gives reasonable results 
as long as the interference between reaction and the Bardeen-Herring back correlation 
is absent. 

The equation can be further simplified by ignoring the memory in the diffusion kernel. 
In this approximation, 
the survival probability and the recombination probability for localized reactions 
can be calculated respectively from 
Eq. (\ref{suv_invFourier_sol}) and Eq. (\ref{RPl})  
by introducing the correlation factor into the hopping frequency,  
\begin{align}
\gamma_B \rightarrow  \gamma_B f_c,  
 \label{suve25}
\end{align}
where the correlation factor $f_c=\hat{f}_c(s=0)$ is given by  
\begin{align}
f_c = \frac{1-\mu}{\displaystyle 1 - \mu  
\frac{\gamma_w+\gamma_B(1-3 c)}{\gamma_w+\gamma_B(1-c)}}  , 
\label{suve26}
\end{align}
and $\mu=\mu(s=0)$ is known for some lattices. 
When the hopping frequency are the same for the inert particles and the reactant, 
$\gamma_w=\gamma_B$, 
the value of $\mu$ is $0.20984$ and $1-(2/\pi)=0.363$ for the cubic and the square lattice, respectively.

The summary of this section is as follows. 
We have studied the influence of back-jump correlations on the survival probability of a geminate pair 
when the initial distribution of inert particles is uniform. 
When the mobile reactant hops to a vacant site, 
the reactant tends to jump back to its previously occupied empty site   
(the Bardeen-Herring back correlation). 
In this way, the reactant motion is highly correlated with the time dependent arrangements of the inert particles. 
The back-jump correlations interfere with reaction. 
In principle, the interference can be 
taken  into account by Eqs. (\ref{suve9}) and (\ref{suve10}) with Eq. (\ref{suve12}). 
However, in practice these equations are hard to solve. 
If the interference is ignored, 
the influence of back-jump correlations is taken into account by Eqs. (\ref{suve14})-(\ref{suve16_1}) with 
Eq. (\ref{suve18_B5}). 
By introducing further simplification of ignoring the memory effect, 
we obtain Eq. (\ref{suv5}) with substitution given by Eq. (\ref{suve25}). 
In the next section, these results will be compared with those 
obtained by numerical simulations.

\section{Comparison to Simulation Results}
\vspace{0.5cm}

\subsection{Simulation method}
In order to see the interference of reaction with the correlated diffusion, 
we perform Monte-Carlo numerical simulations. 
We numerically obtain the probability of geminate reaction in the presence of site-blocking effects using a kinetic Monte Carlo method. The simulation is carried out on the simple cubic lattice. One reactant is placed at the lattice site 
$(0,0,0)$ and assumed to be immobile. The other reactant is initially placed at  $(j,0,0)$, where $j$ is an integer and the lattice constant is unity. Inert particles are randomly generated at lattice sites within the box  
$(0,L-1)^3$, where $L$ is the box length. The number of inert particles, $N$, is related to their concentration by $c=N/L^3$. Each lattice site may accommodate only one inert particle or the mobile reactant. We assume that the inert particles belonging to the box $(0,L-1)^3$ are periodically replicated in three dimensions, so that the simulation volume is effectively unlimited. What should be noted is that the spatial periodicity is assumed only for the distribution of inert particles, and not for the reactants themselves.
During the simulation, both the inert particles and the mobile reactant may perform hops to neighboring lattice sites. A hop is allowed only when the destination site is not occupied by another inert particle or the mobile reactant. However, it is allowed for both types of simulated particles to jump to $(0,0,0)$. If the mobile reactant is staying at $(0,0,0)$, its reaction with the other reactant is possible. The procedure of selecting the event that actually takes place at a given simulation step is as follows. First, we determine all possible hops for the mobile reactant and inert particles. Denote the numbers of such hops as $K$ and $K_{\rm in}$, respectively. If the mobile reactant is staying at a site other than $(0,0,0)$, the total rate of all possible events is calculated as $k_{\rm tot}=K \gamma_B + K_{\rm in} \gamma_w$. Otherwise, the total rate includes the rate of reaction and is calculated as  $k_{\rm tot}=K \gamma_B + K_{\rm in} \gamma_w+k_0$. Now, we determine which event will actually take place. This is decided at random, taking the ratio of the rate of each possible event to the total rate $k_{\rm tot}$ as the event probability. The above procedure is repeated 
until either a reaction occurs or the mobile reactant separates to a large distance $r_{\rm max}$ from $(0,0,0)$. By repeating the simulation for a large number (at least $2\times10^4$) of independent runs, we can obtain the reaction probability.
The accuracy of the simulation results depends on two parameters: $L$ and $r_{\rm max}$. They should be taken as large as possible within the practical limits imposed by the available computational time (the demand on computer time is especially high at large concentrations of inert particles). In the production runs of the simulation, we assumed $L=10$ and 
$r_{\rm max} =30$. From test calculations carried out also for other values of these parameters, we found no significant effect of $L$ on the obtained results. However, a weak dependence of the reaction probability on the value of $r_{\rm max}$ could be observed. For example, the reaction probability obtained for $j=3$ and $c=0$ with  $r_{\rm max}=60$ was about 2\% higher than that calculated with  $r_{\rm max}=30$.

\subsection{Simulation results}

We investigate quantitatively the effects of the factors ignored in deriving simple result, Eq. (\ref{suv5}), by comparison to the 
more rigorous theoretical results and the simulation results. 

One of the factors ignored is 
the Bardeen-Herring back correlation. 
The Bardeen-Herring back correlation is described by 
the four-point correlation function given in Appendix \ref{appA}.  
The Bardeen-Herring back correlation is taken into account fully by Eqs. (\ref{suve14})-(\ref{suve16_1}) and partly by 
Eq. (\ref{suv5}) with Eq.(\ref{suve25}). 
Equation (\ref{suve25}) is obtained from Eqs. (\ref{suve14})-(\ref{suve16_1}) by taking 
the limit of small wave length, $k \rightarrow 0$ and ignoring the memory effect. 
Equation (\ref{suv5}) with Eq. (\ref{suve25}) is much simpler than Eqs. (\ref{suve14})-(\ref{suve16_1}). 
The numerical way to solve 
Eqs. (\ref{suve14})-(\ref{suve16_1}) with the additional set of equations are given below Eq. (\ref{suve18_B5}) of  Appendix \ref{appA}.  

Another factor ignored is the effect of the interference between reaction and the Bardeen-Herring back correlation. 
The interference is taken into account in the results of numerical simulations but 
is ignored in any theoretical results including 
the most sophisticated one given by the solution of Eqs. (\ref{suve14})-(\ref{suve16_1}). 

\begin{figure}[htbp]
\centerline{\includegraphics[width=0.55\columnwidth]{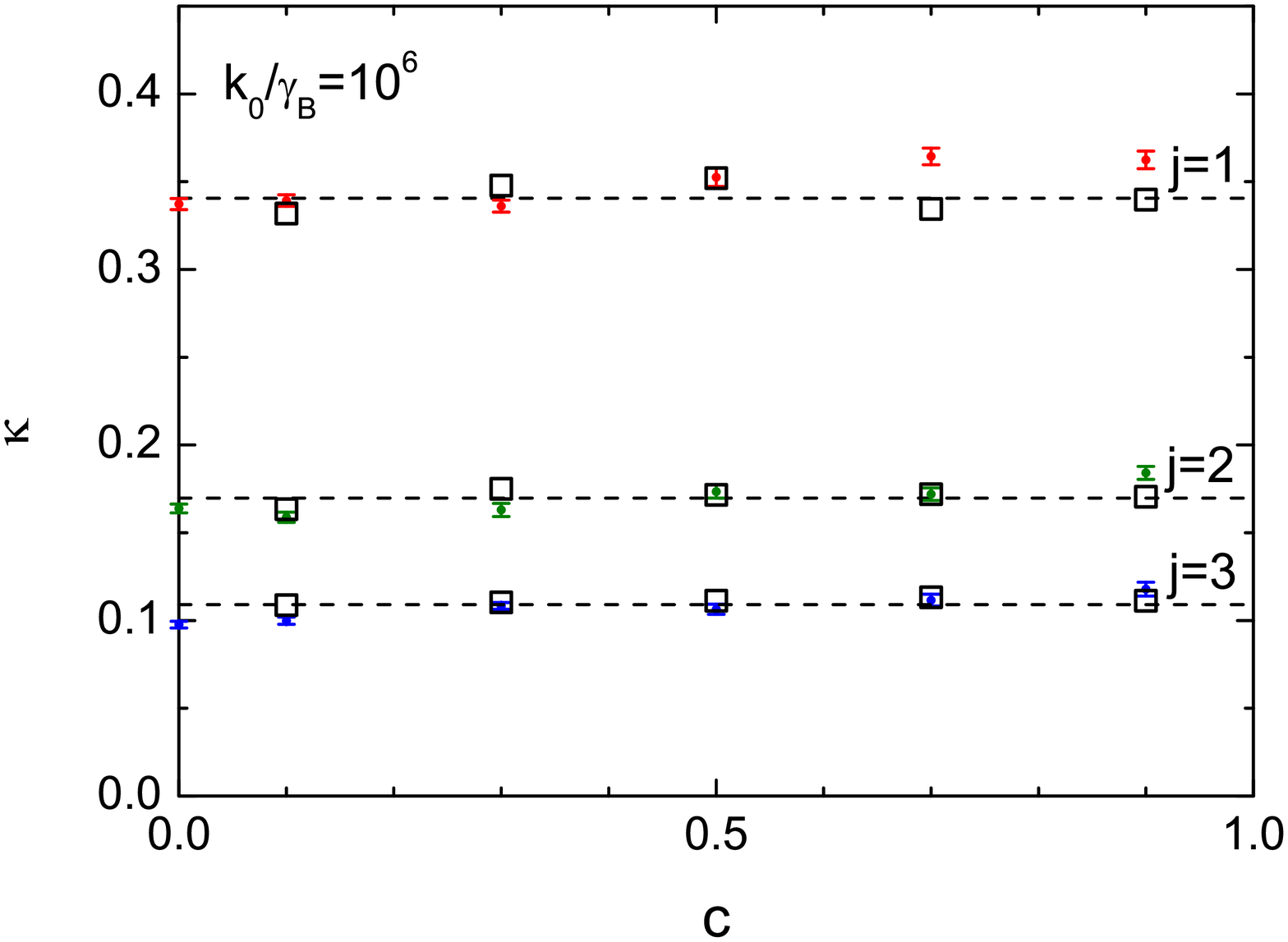}}
\caption{
Recombination probability of a geminate pair against concentration of inert particles, $c$, for 
$k_0 \rightarrow \infty$ (In the simulation, $k_0/\gamma_B=10^6$) and $\gamma_w/\gamma_B=1$.  
$j$ indicates the initial separation of the geminate pair. 
Dots with error bars indicate the simulation results. 
Squares represent the numerical solutions of Eqs. (\ref{suve14})-(\ref{suve16_1}) with Eq. (\ref{suve18_B5}). 
Dashed lines indicate the mean field results of Eq. (\ref{RPl}). 
The results of Eq. (\ref{RPl}) with the substitution given by Eq. (\ref{suve25}) 
practically coincide with those of Eq. (\ref{RPl}). 
}
\label{fig:homogeneous_inf}
\end{figure} 
\begin{figure}[htbp]
\centerline{\includegraphics[width=0.5\columnwidth]{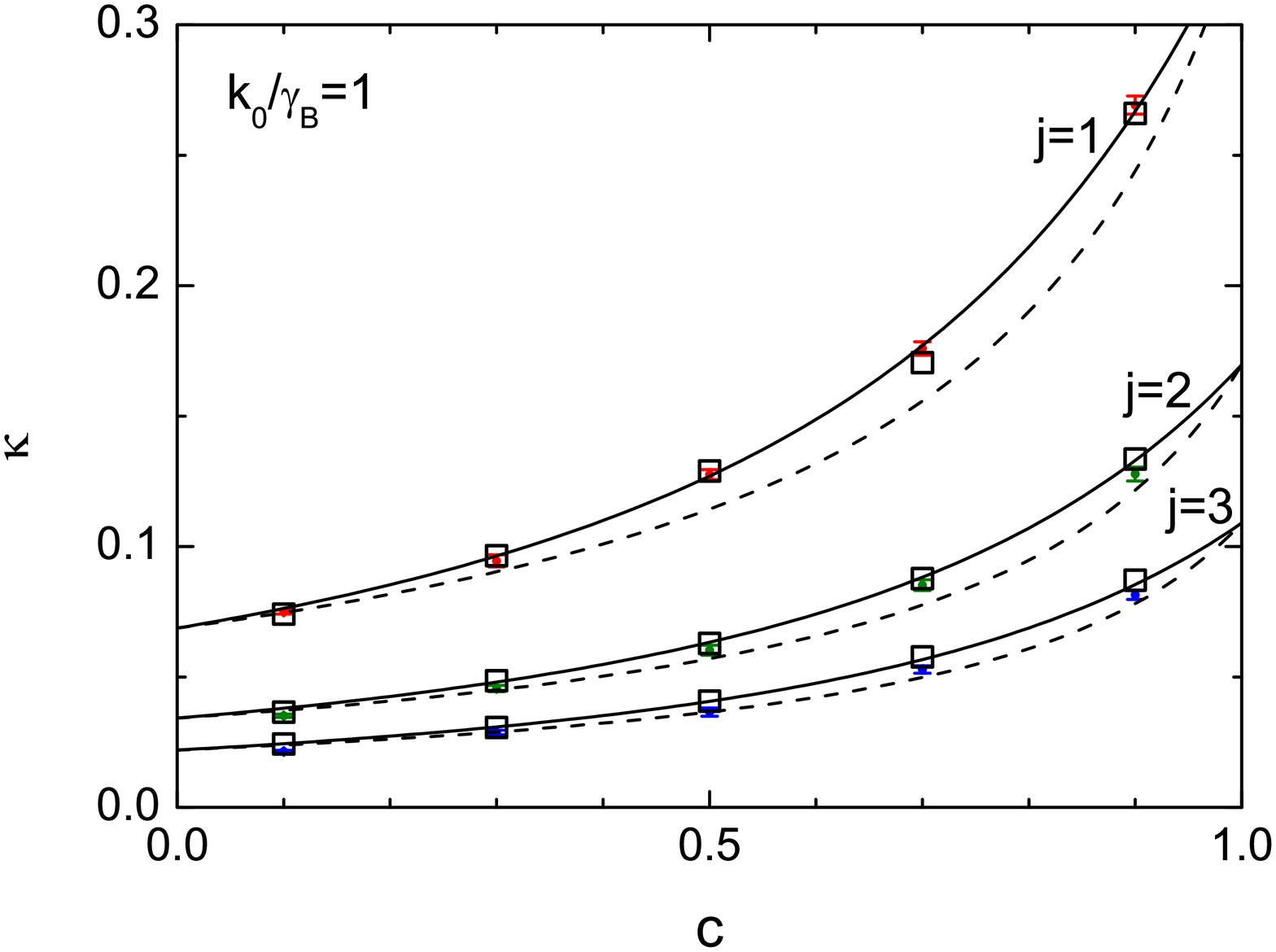}}
(a)
\centerline{\includegraphics[width=0.5\columnwidth]{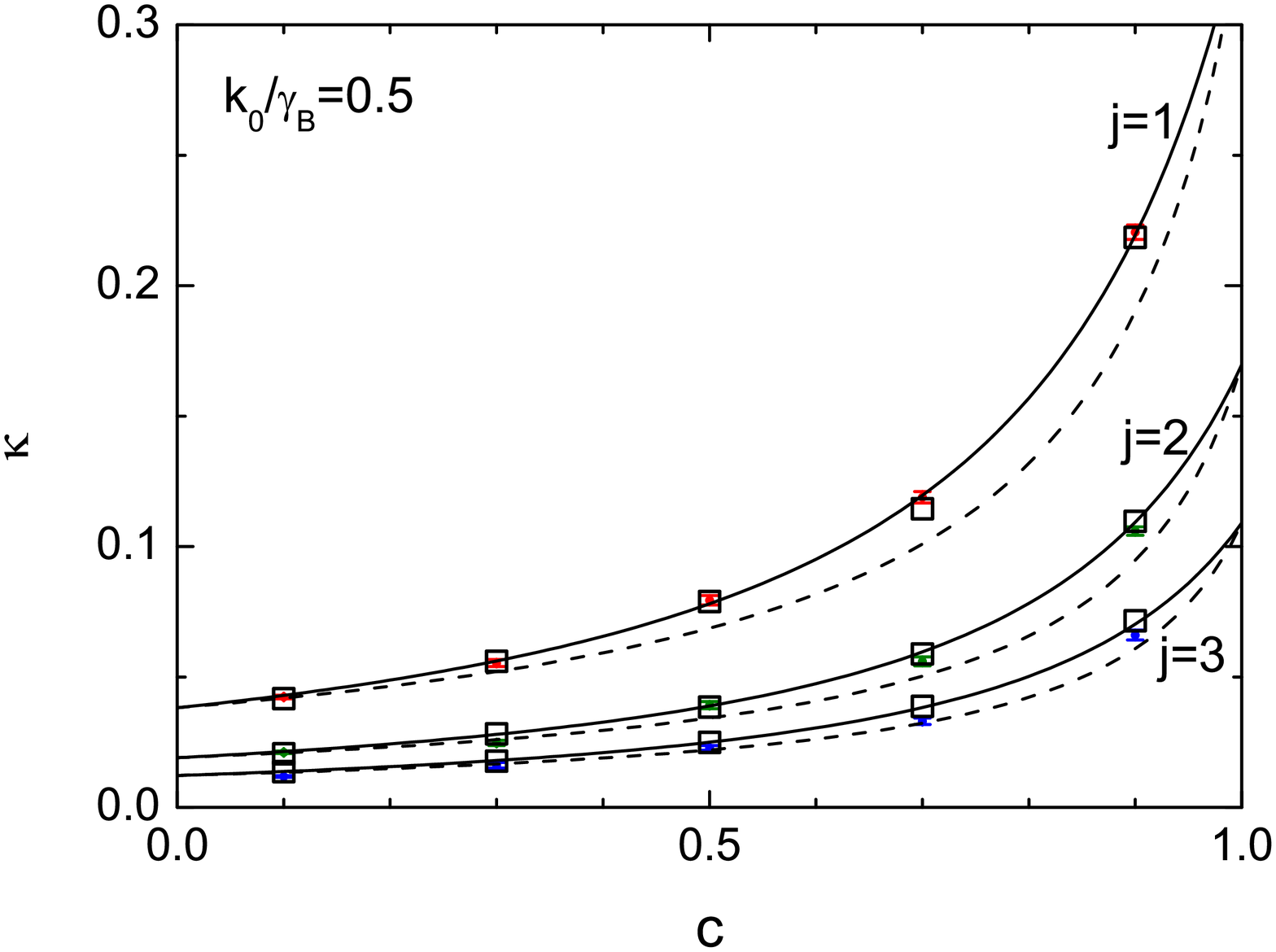}}
(b)
\caption{
Recombination probability of a geminate pair against the concentration of the inert particles, $c$, when $\gamma_w/\gamma_B=1$. 
(a)  $k_0/\gamma_B=1.0$ and (b) $k_0/\gamma_B=0.5$. 
$j$ indicates the initial separation of the geminate pair. 
Dots with error bars indicate the simulation results. 
Squares represent the numerical solutions of Eqs. (\ref{suve14})-(\ref{suve16_1}) with Eq. (\ref{suve18_B5}). 
The results of Eq. (\ref{RPl}) with the substitution given by Eq. (\ref{suve25}) are indicated by the solid lines. 
Dashed lines indicate the mean field results of Eq. (\ref{RPl}). 
}
\label{fig:homogeneous_1half}
\end{figure} 

The recombination probability in the completely diffusion controlled limit, $k_0 \rightarrow \infty$, is shown in 
Fig. \ref{fig:homogeneous_inf}. 
The simulation results are compared to the numerical solutions of 
Eqs. (\ref{suve14})-(\ref{suve16_1}). 
In the theoretical results the influence of reaction on the four-point correlation function is ignored.  
As shown in Fig. \ref{fig:homogeneous_inf}, 
the influence of reaction on the four-point correlation function is 
relatively small for all concentrations of inert particles 
as long as overall yield (recombination probability) is concerned. 
However, Fig. \ref{fig:homogeneous_inf} shows small deviation at 
high concentration of inert particles for the initial separation of $j=1$. 
The deviation is not an effect of 
statistical error of the simulation. 
We will discuss this point later 
when we study transient decay of the survival probability. 

In the completely diffusion-controlled limit, $k_0 \rightarrow \infty$, 
the recombination probability obtained 
from Eq. (\ref{RPl}) with 
the substitution given by 
Eq. (\ref{suve25}) is independent of the concentration of the inert particles. 
This result is not a rigorous relation. 
It is obtained by the oversimplification of taking the limit of small wave length, $k \rightarrow 0$, and 
ignoring  interference between reaction and the Bardeen-Herring back correlation. 
However, the difference between the result of the simplified equation, Eq. (\ref{RPl}) with Eq. (\ref{suve25}), and that obtained  
without taking the limit of long-wave length, 
Eqs. (\ref{suve14})-(\ref{suve16_1}), is also very small. 
Figure \ref{fig:homogeneous_inf} indicates that 
the simplified approach which leads to Eq. (\ref{RPl}) with Eq. (\ref{suve25}) 
is justified for the calculation of the recombination probability in the diffusion-controlled limit.

The recombination probability in the case of finite reactivity is shown in Fig. \ref{fig:homogeneous_1half}. 
The influence of reaction on the four-point correlation function can be seen as the difference between 
the simulation results and the numerical solutions of Eqs. (\ref{suve14})-(\ref{suve16_1}). 
In the reaction-controlled limit, 
the difference is very small for all concentrations of the inert particles 
regardless of the initial distance of a geminate pair. 
The difference between the simplified results obtained 
Eq. (\ref{RPl}) with Eq. (\ref{suve25}) and the solutions  
of Eqs. (\ref{suve14})-(\ref{suve16_1})   
is again negligibly small.  

\begin{figure}[htbp]
\centerline{\includegraphics[width=0.5\columnwidth]{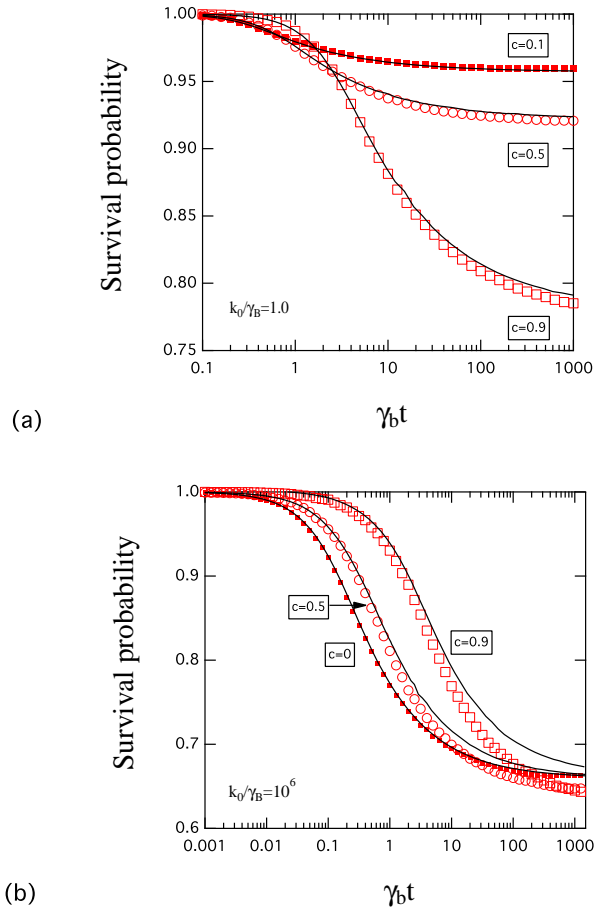}}
\caption{ 
Survival probability as a function of $\gamma_B t$ for the initial separation of $j=1$. 
(a)  $k_0/\gamma_B=1.0$ and (b) $k_0/\gamma_B=10^6$. 
Concentrations of inert particles are shown in figures. 
Symbols represent the simulation results. 
Lines indicate the theoretical results of Eqs. (\ref{suv_invFourier_sol})-(\ref{latticeU}) with Eq. (\ref{psiBs}). 
}
\label{fig:homogeneous_kinetics}
\end{figure} 
We also compare transient decay of the survival probability obtained from simulations with 
theoretical results of Eqs. (\ref{suv_invFourier_sol})-(\ref{latticeU}) with Eq. (\ref{psiBs}) 
to study the effect of interference between reaction and the Bardeen-Herring back correlation.
Figure \ref{fig:homogeneous_kinetics} (a) shows that 
the theoretical results reproduce simulation results in the reaction-controlled limit ($k_0/\gamma_B=1.0$) 
over the whole time range regardless of the concentration of inert particles. 
As we have theoretically shown in the previous section, 
the interference between reaction and the Bardeen-Herring back correlation changes the diffusion term through 
4-point correlation functions, leaving the reaction term unaltered. 
The interference could be dominated in the diffusion-controlled limit but should be small in the reaction-controlled limit. 
Figure \ref{fig:homogeneous_kinetics} (a) indicates that the approximation 
used for finite $k$ in 
Eqs. (\ref{suv_invFourier_sol})-(\ref{latticeU}) and Eq. (\ref{psiBs}) is valid as long as 
interference between reaction and the Bardeen-Herring back correlation can be ignored. 
Figure \ref{fig:homogeneous_kinetics} (b) shows that in the diffusion-controlled limit 
the theoretical results agree with the simulation results 
in the absence of inert particles. 
However, when the 
concentration of inert particles is high ($c>0.5$),  
the survival probability obtained from simulation decays faster compared to the theoretical results. 
The acceleration of the decay should be attributed to the influence of interference between reaction and 
the Bardeen-Herring back correlation. 

As the conclusion of this section, 
we point out that 
the interference between reaction and the Bardeen-Herring back correlation is most pronounced in the transient decay of the survival probability 
in the diffusion-controlled limit at high concentration of inert particles. 
As long as the reaction yield is concerned, 
the 
mean field results given by Eq. (\ref{suv5}) reproduce the simulation results regardless of reaction strength and 
the concentration of the inert particles when the density of the inert particles is uniform. 
The result can be improved by introducing substitution given by Eq. (\ref{suve25}) into Eq. (\ref{suv5}) to take into account 
the Bardeen-Herring back correlation. 
Generalization of Eq. (\ref{suv5}) to the case of continuous diffusion and non-uniform distribution of inert particles 
is shown in the subsequent sections.

\section{Inhomogeneous distribution of inert particles}
\vspace{0.5cm}

So far, we have assumed the homogeneous distribution of inert particles. 
Recently, the influence of inhomogeneous distributions of inert particles 
is taken into account to study catalytic surface reactions, in particular focusing on  
reaction front structures. \cite{McEwen,Zhdanov,Tammaro,Liu,Evans92,Tammaro95} 
In this section, we study geminate reactions under the inhomogeneous distribution of inert particles. 
Since we are not able to solve the lattice model for nonuniform initial distribution of inert particles 
with the same rigor as that under uniform distribution of inert particles, 
we study the results in 
the continuous limit by using the mean field approximation. 
We consider the pair distribution $p(\vec{r},t)$ of finding a pair of reactants at the separation $\vec{r}$ at time $t$.  
As shown in Appendix \ref{appB}, the lattice model considered in this paper leads to 
\begin{align}
\frac{\partial}{\partial t} p \left(r, t \right)  =  \vec{\nabla}\cdot D_B
\left[ (1-c_v \left(\vec{r}, t \right)) \vec{\nabla} p \left(r, t \right) 
+ p \left(r, t \right)\vec{\nabla} c_v \left(\vec{r}, t \right)    \right]
- k \left( r \right) p \left(r, t \right) 
, 
\label{suv_cont_1}
\end{align}
in the continuous limit, 
where 
the diffusion constant is defined by  
$D_B=b^2 \gamma_B$ and 
the concentration of inert particles is denoted by $c_v(\vec{r},t)$. 
The first term on the right-hand side includes the 
diffusion term influenced by the concentration gradient of inert particles  
\cite{McEwen,Zhdanov,Tammaro,Liu,Evans92,Tammaro95} 
and the drift term is induced by the spurious potential defined by, 
\begin{align}
U= - k_B T \ln \left[ 1 - c_v  \left(r, t \right) \right].   
\label{potential}
\end{align}
Equation (\ref{suv_cont_1}) can be rewritten in terms of the potential as 
\begin{align}
\frac{\partial}{\partial t} p \left(r, t \right)  = \vec{\nabla} \cdot D_B (1-c_v \left(\vec{r}, t \right)) 
\left[ \vec{\nabla} p \left(r, t \right) 
+p \left(r, t \right)  \frac{\vec{\nabla} U}{k_B T} 
\right]  
- k \left( r \right) p \left(r, t \right) .
\label{suv_cont_2}
\end{align}
The perfectly reflecting boundary condition at $R$ is imposed to express that  
the reactants cannot penetrate each other.

We calculate the escape probability on the basis of Eq. (\ref{suv_cont_2}). 
We consider the case that the density of the inert particles is stationary 
and inhomogeneous. 
Both the intrinsic reaction rate and the density of the inert particles are assumed to be isotropic. 
The equation for the survival probability is obtained from Eq. (\ref{suv_cont_2}) 
by introducing the adjoint operator as, \cite{Tachiya78,Sano,Szabo}
\begin{align}
\frac{\partial}{\partial t}  S \left(r, t \right) = \vec{\nabla} \cdot D_B (1-c_v \left(r \right)) \vec{\nabla} S \left(r, t \right) 
-D_B \left[ \vec{\nabla} c_v\left(r \right) \right] \cdot \vec{\nabla} S \left(r, t \right) - k \left( r \right) S \left(r, t \right) ,  
\label{eq:Adjoint}
\end{align}
where $\vec{\nabla}$ in the square brackets operates only on $c_v \left(r \right)$. 
The perfectly reflecting boundary condition at $r=R$ is represented by, 
\begin{align}
\left. \frac{\partial}{\partial r} S \left(r, t \right) \right|_{r=R} =0. 
\label{eq:suv_pr}
\end{align}
When the reaction takes place at the reaction radius, $R$, with the intrinsic rate, $k_0$, \cite{Collins} 
the equation for the escape probability defined by $\varphi(r)= \lim_{t \rightarrow \infty} S \left(r, t \right)$ 
in $d$-dimension satisfies, 
\begin{align}
\frac{1}{r^{d-1}} \exp\left( \frac{U}{k_B T} \right) 
\frac{\partial}{\partial r} D_B (1-c_v \left(r \right)) r^{d-1}\exp\left( -\frac{U}{k_B T} \right) 
\frac{\partial}{\partial r} \varphi(r) =0, 
\label{eq:esc}
\end{align}
using the potential defined by Eq. (\ref{potential}). 
The boundary conditions are given by $\lim_{r\rightarrow \infty} \varphi(r)=1$ and 
\begin{align}
\left. S_d D_B (1-c_v \left(R \right)) \frac{\partial}{\partial r} \varphi \left(r \right) \right|_{r=R} =k_0 \varphi (R),  
\label{eq:suv_par}
\end{align}
where the surface area of the $d$-dimensional sphere is given by 
$S_d = d \pi^{d/2}/\Gamma((d/2)+1)$.  
For 2 and 3 dimensions, 
$S_2= 2 \pi R$ and $S_3= 4 \pi R^2$. 
The solution of Eq. (\ref{eq:esc}) subject to the above mentioned boundary conditions is obtained as, 
\begin{align}
\varphi (r) = \frac{\displaystyle 
\int_R^r d r_1 \frac{1}{D_B (1-c_v \left(r_1 \right))^2 r_1^{d-1}}
+ \frac{S_d}{k_0} \frac{1}{ (1-c_v \left(R \right)) R^{d-1}
}}
{\displaystyle 
\int_R^\infty d r_1 \frac{1}{D_B (1-c_v \left(r_1 \right))^2 r_1^{d-1}} 
+ \frac{S_d}{k_0} \frac{1}{ (1-c_v \left(R \right)) R^{d-1}
}} . 
\label{eq:escsol}
\end{align}
In the limit of perfectly absorbing boundary condition, 
the escape probability simplifies into,  
\begin{align}
\varphi (r) = 
\int_R^r d r_1 \frac{1}{(1-c_v \left(r_1 \right))^2 r_1^{d-1}}/
\int_R^\infty d r_1 \frac{1}{(1-c_v \left(r_1 \right))^2 r_1^{d-1}} . 
\label{eq:escsol_infty}
\end{align}

According to Eq. (\ref{eq:escsol_infty}), 
the escape probability is independent of the density of the inert particles when 
the density is homogeneous. 
This result is consistent with that obtained in the lattice system in Sec. IV. 
On the other hand, when the density of the inert particles is inhomogeneous, 
the escape probability in general depends on the density of the inert particles. 
The escape probability is lower than 
that for the homogeneous density of inert particles if $c_v \left(r \right)$ has a positive slope. 
The recombination reaction can be assisted by a positive density gradient of the inert particles. 
On the other hand, the recombination can be hindered by a negative density gradient of the inert particles.

\section{Conclusions}
\vspace{0.5cm}

In this paper,  
the time evolution equations for 
the survival probability of a geminate pair under the presence of many inert particles 
are derived and the results are compared to the simulation results. 
If we ignore correlations higher than two-point correlations, 
Eq. (\ref{suv5}) is derived. 
In this lowest order approximation, 
the influence of inert particles is described by 
using the mean field expression of the tracer-diffusion constant in the reaction-diffusion equation. 

In the lowest order approximation, 
the so-called Bardeen-Herring back correlation is not taken into account. 
The Bardeen-Herring back correlation is the tendency of the 
preferred jump of the diffusing particle back to the previously occupied empty site. 
We have shown that 
the reaction interferes with 
the Bardeen-Herring back correlation. 
In the reaction-diffusion equation, 
the transition operator describing diffusion is influenced by the reaction strength 
while leaving the reaction term unaltered. 
If the interference between the reaction and the Bardeen-Herring back correlation is taken into account,  
the reaction-diffusion equation becomes very complicated and cannot be solved analytically. 
By taking into account the Bardeen-Herring back correlation but ignoring the interference between reaction and the Bardeen-Herring back correlation, 
we obtain the reaction-diffusion equation given in terms of 
the improved expression of the tracer-diffusion constant. 
The influence of the excluded volume interactions 
is taken into account solely by the tracer-diffusion constant. 
The tracer-diffusion constant decreases by increasing the concentration of the inert particles 
since the diffusive motion of reactive species is hindered by the presence of the inert particles.

By comparison of the theoretical results with the results of numerical simulations 
the interference between reaction and the Bardeen-Herring back correlation is shown to be small 
as long as overall reaction yield is concerned. 
The interference also influences 
the transient decay of the survival probability of a geminate pair 
in the diffusion-controlled limit. 
When the initial concentration of the inert particles is high, 
we show that interference between reaction and the Bardeen-Herring back correlation accelerates 
the transient decay of the survival probability in the diffusion-controlled limit.

Recently, Schmit {\it et al.} studied the reaction in microfluid by the Monte-Carlo simulation. \cite{Schmit}
The simulation results are well approximated by the survival probability of a pair of reactants 
obtained from the reaction-diffusion equation similar to Eq. (\ref{suv5}): 
it is suggested to use 
the reaction-diffusion equation in which 
the mutual diffusion coefficient is substituted by the effective tracer-diffusion coefficient 
of a tagged particle in a sea of inert particles. \cite{Schmit} 
They calculated the mean first passage time instead of the survival probability. 
Since the mean first passage time can be given in terms of the inverse of the overall reaction rate, 
their results are consistent with our conclusion,  
although  the slightly different expression of the effective tracer-diffusion coefficient is 
used in their equation. 
According to our work, the approximation would fail in the transient decay of the survival probability 
in the diffusion controlled limit at high concentration of inert particles.  

The above conclusions are obtained by assuming the homogeneous distribution of the inert particles. 
We have also formulated a way to obtain the survival probability of a geminate pair 
when the initial distribution of inert particles is inhomogeneous. 
The reaction yield is increased when the reaction proceeds in the presence of 
a positive density gradient of the inert particles which inhibits the escape of reactants. 
The effect can be interpreted as a cage effect. 
Although 
we need further investigation for the kinetics of the survival probability 
in the presence of the density gradient of the inert particles, 
our results on the escape probability indicate that the crowding promotes reactions when  
 the density of the inert particles increases with the distance from the location of the immobile reactant.

 Recently, the cage effect of crowding by inert particles was introduced by Kim and Yethiraj 
 to interpret the increase of association rate constant with increasing the concentration of 
 inert particles obtained by Brownian dynamic simulations when the intrinsic reaction rate constant was small. \cite{Yethiraj}
 The cage effect in this paper is similar to that introduced by them 
 in the sense that the presence of inert particles surrounding a reactant pair promotes the reaction 
 but the mechanism is slightly different. 
 In our case, the concentration gradient toward one of the reactant pair assists the recombination reaction, 
 while the association rate in their case increases by the increase of the recollision probability due to the 
 high concentration of inert particles rather than the concentration gradient.


\appendix
\section{Diffusion kernel}
\label{appA}

By introducing the operator identity, 
\begin{align}
\frac{1}{s- \tilde{L}_{\rm rw0}-  \tilde{L}_{\rm rc}} 
= \frac{1}{s} \left( 
1+ \frac{1}{s- \tilde{L}_{\rm rw0}-  \tilde{L}_{\rm rc}} 
\left( \tilde{L}_{\rm rw0}+  \tilde{L}_{\rm rc} \right)
\right) ,  
\label{suve11}
\end{align}
into Eq. (\ref{suve7}), 
and from the definition given by Eq. (\ref{suve7}) 
we derive, 
\begin{multline}
s G\left(\vec{n} \bullet ,\vec{n}+\vec{b}_r \circ |  \vec{m} \bullet, \vec{r} \circ  ,s \right)
- \delta_{\vec{r},\vec{b}_r} \delta_{\vec{n},\vec{m}} = 
\sum_\alpha \left[ 
\gamma_w G\left(\vec{n} \bullet ,\vec{n}+\vec{b}_r \circ |  \vec{m} \bullet, \vec{r} +\vec{b}_\alpha \circ  ,s \right)
\right. \\ \left.
+ 
\gamma_B (1-c) G\left(\vec{n} \bullet ,\vec{n}+\vec{b}_r \circ |  \vec{m}-\vec{b}_\alpha \bullet, \vec{r}+\vec{b}_\alpha \circ  ,s \right)
-\gamma_t G\left(\vec{n} \bullet ,\vec{n}+\vec{b}_r \circ |  \vec{m} \bullet, \vec{r} \circ  ,s \right)
\right]
\\ 
- \delta_{\vec{r},\vec{0}} 
\sum_\alpha \left[ 
\gamma_w G\left(\vec{n} \bullet ,\vec{n}+\vec{b}_r \circ |  \vec{m} \bullet, \vec{b}_\alpha \circ  ,s \right)
+ 
\gamma_B (1-c) G\left(\vec{n} \bullet ,\vec{n}+\vec{b}_r \circ |  \vec{m} -\vec{b}_\alpha \bullet, \vec{b}_\alpha \circ  ,s \right)
\right] \\
+ \sum_\alpha \delta_{\vec{r},\vec{b}_\alpha} 
\left[ 
\gamma_B c G\left(\vec{n} \bullet ,\vec{n}+\vec{b}_r \circ |  \vec{m}+\vec{b}_\alpha \bullet, -\vec{b}_\alpha \circ  ,s \right)
+ 
\left(\gamma_t- \gamma_B c \right) G\left(\vec{n} \bullet ,\vec{n}+\vec{b}_r \circ |  \vec{m} \bullet, \vec{b}_\alpha \circ  ,s \right)
 \right] \\ 
 - 
k \left( \vec{m} \right) G\left(\vec{n} \bullet ,\vec{n}+\vec{b}_r \circ |  \vec{m} \bullet, \vec{r} \circ  ,s \right) ,
\label{suve12}
\end{multline}
where 
\begin{align}
\gamma_t = \gamma_w+ \gamma_B (1-c) . 
\label{gammt}
\end{align} 
In the presence of $k \left( \vec{m} \right)$, 
$G\left(\vec{n} \bullet ,\vec{n}+\vec{b}_r \circ |  \vec{m} \bullet, \vec{r} \circ  ,s \right)$ does not satisfy 
the condition of translational invariance of $\vec{m}$ against $\vec{n}$.

If we ignore $k \left( \vec{m} \right)$ in Eq. (\ref{suve12}), 
the translational invariance is satisfied. 
The equation for $G^{(T)} \left(\vec{\ell} , {\vec{b}_r}\,|  \vec{r} ,s \right)$ introduced in Eq. (\ref{suve13}) is given by, 
\begin{multline}
s G^{(T)} \left( \vec{\ell} , {\vec{b}_r}\,| \vec{r} ,s \right) - \delta_{\vec{r},\vec{b}_r} \delta_{\vec{\ell},\vec{0}} = 
\sum_\alpha \left[ 
\gamma_w G^{(T)} \left( \vec{\ell} , {\vec{b}_r}\,| \vec{r} +\vec{b}_\alpha,s \right)+ 
\gamma_B (1-c) G^{(T)} \left( \vec{\ell}-\vec{b}_\alpha  , {\vec{b}_r}\,| \vec{r} +\vec{b}_\alpha,s \right) 
\right. \\ \left.
-\gamma_t G^{(T)} \left( \vec{\ell}  , {\vec{b}_r}\,| \vec{r} ,s \right)
\right]
- \delta_{\vec{r},\vec{0}} 
\sum_\alpha \left[ 
\gamma_w G^{(T)} \left( \vec{\ell}  , {\vec{b}_r}\,| \vec{b}_\alpha,s \right)+ 
\gamma_B (1-c) G^{(T)} \left( \vec{\ell}-\vec{b}_\alpha  , {\vec{b}_r}\,| \vec{b}_\alpha,s \right) 
\right] \\
+ \sum_\alpha \delta_{\vec{r},\vec{b}_\alpha} 
\left[ 
\gamma_B c G^{(T)} \left( \vec{\ell}+\vec{b}_\alpha  , {\vec{b}_r}\,| -\vec{b}_\alpha,s \right)+ 
\left(\gamma_t- \gamma_B c \right) 
G^{(T)} \left( \vec{\ell}  , {\vec{b}_r}\,| \vec{b}_\alpha,s \right) \right] . 
\label{suve17}
\end{multline}
By applying Fourier transformation, 
\begin{align}
g \left( \vec{k}  , {\vec{b}_r}\,| \vec{h},s \right) &= \sum_{\vec{\ell}} \sum_{\vec{r}}
\exp\left[i \left( \vec{k} \cdot \vec{\ell} + \vec{h} \cdot \vec{r} \right) \right]
G^{(T)} \left( \vec{\ell}  , {\vec{b}_r}\,| \vec{r},s \right)
\label{suve17_B1}
\\
\tilde{G} \left( \vec{k}  , {\vec{b}_r}\,| \vec{r},s \right)
&= 
\sum_{\vec{\ell}} 
\exp\left[i  \vec{k} \cdot \vec{\ell}  \right]
G^{(T)} \left( \vec{\ell}  , {\vec{b}_r}\,| \vec{r},s \right)
\label{suve17_B1_1}
\end{align}
we obtain, 
\begin{multline}
s g \left( \vec{k}  , {\vec{b}_r}\,| \vec{h} ,s \right)   = \exp \left( i \vec{h} \cdot \vec{b}_r \right) +
\sum_\alpha \left[ 
\omega_t \left( \vec{b}_\alpha \right)  \exp \left(- i \vec{h} \cdot \vec{b}_\alpha \right) 
-\gamma_t 
\right] 
g \left( \vec{k}  , {\vec{b}_r}\,| \vec{h},s \right)
 \\
+ \sum_\alpha \left[ \exp \left(- i \vec{h} \cdot \vec{b}_\alpha \right)  
\omega_B \left(  \vec{b}_\alpha \right)c + 
\gamma_s \exp \left( i \vec{h} \cdot \vec{b}_\alpha \right) -\omega_t \left( \vec{b}_\alpha \right) \right] 
\tilde{G} \left( \vec{k}  , {\vec{b}_r}\,| \vec{b}_\alpha,s \right) ,
\label{suve17_B2}
\end{multline}
where we define, 
\begin{align}
\gamma_s &= \gamma_t - \gamma_B c , \\
\omega_B\left(\vec{r} \right)= \gamma_B \exp \left(i \vec{k} \cdot \vec{r} \right),  &\mbox{ and }
\omega_t \left(\vec{r} \right)= \gamma_w+\gamma_B(1-c) \exp \left(i \vec{k} \cdot \vec{r} \right) .
\label{suve18_B3}
\end{align}
It is convenient to introduce the function, 
\begin{align}
Q(\vec{k}, \vec{r}, s)= 
\frac{1}{(2\pi)^d} \int \cdots \int_{-\pi}^{\pi} d^d \vec{h}
\frac{\exp \left( - i \vec{h} \cdot \vec{r} \right)}{s + \Gamma_t - \Omega_t (\vec{k},\vec{h})} , 
\label{suve18_B4}
\end{align}
where $\Gamma_t=2d \gamma_t$ and $\Omega_t (\vec{k},\vec{h})$ is defined by, 
\begin{align}
\Omega_t (\vec{k},\vec{h}) = 2d \left[ \gamma_w \lambda(\vec{h}) + \gamma_B (1-c) \lambda(\vec{k} - \vec{h})
\right] .
\label{suve18_B4_1}
\end{align}
A closed set of equations can be obtained from Eq. (\ref{suve17_B2}) as,  
\begin{multline}
\tilde{G} \left( \vec{k}  , {\vec{b}_r}\,| \vec{b}_q,s \right)= Q( \vec{k},\vec{b}_q- \vec{b}_r, s) +
\sum_\alpha \left[ \gamma_s Q( \vec{k},\vec{b}_q-\vec{b}_\alpha, s)  +
\omega_B \left(  \vec{b}_\alpha \right)c  
Q( \vec{k},\vec{b}_q+\vec{b}_\alpha, s) - \right.
\\ \left.
\omega_t \left( \vec{b}_\alpha \right) Q( \vec{k},\vec{b}_q, s) \right] 
\tilde{G} \left( \vec{k}  , {\vec{b}_r}\,| \vec{b}_\alpha,s \right) . 
\label{suve18_B5}
\end{multline}
The solution is independent of the position of the reactive sink in this approximation. 
By introducing the solution of Eq. (\ref{suve18_B5}) into
Eq. (\ref{suve16}) and using Eq. (\ref{suve14}) and Eq. (\ref{suve15}) 
the survival probability is obtained after numerical  inverse Laplace transformation and inverse Fourier transformation.

An equation for $G^{(a)} \left( {\vec{b}_r}\,|  \vec{r} ,s \right)$ is obtained by applying 
the operator identity, 
\begin{align}
\frac{1}{s- \tilde{L}_{\rm rw0}} 
= \frac{1}{s} \left( 
1+ \frac{1}{s- \tilde{L}_{\rm rw0}} \tilde{L}_{\rm rw0}
\right) ,  
\label{suve21}
\end{align}
as \cite{Nakazato,Suzuki,Okamoto07}
\begin{multline}
s G^{(a)} \left( {\vec{b}_r}\,|  \vec{r} ,s \right) - \delta_{\vec{r},\vec{b}_r} = 
\sum_\alpha \gamma_t \left[ 
G^{(a)} \left({\vec{b}_r}\,| \vec{r} +\vec{b}_\alpha,s \right)
- G^{(a)} \left( {\vec{b}_r}\,|  \vec{r} ,s \right)
\right]
- \delta_{\vec{r},\vec{0}} 
\sum_\alpha 
\gamma_t G^{(a)} \left({\vec{b}_r}\,| \vec{b}_\alpha,s \right)+ 
\\
 \sum_\alpha \delta_{\vec{r},\vec{b}_\alpha} 
\left[ 
\gamma_B c G^{(a)} \left({\vec{b}_r}\,|  -\vec{b}_\alpha,s \right)+ 
\left(\gamma_t- \gamma_B c \right) 
G^{(a)}  \left( {\vec{b}_r}\,|  \vec{b}_\alpha,s \right) \right] . 
\label{suve22}
\end{multline}
The solution depends on a position of an inert particle 
through its relative vector against the initial position of 
a mobile reactant. 

\section{Excluded volume interactions under inhomogeneous distribution of inert particles}
\label{appB}
Since the particle jumps to a vacant site, 
$p(\vec{r},t)$ obeys, 
\begin{align}
\frac{\partial}{\partial t} p(\vec{r},t) = 
\gamma_B \sum_{j} \left( p \left(\vec{r}+\vec{b}_j  \bullet, \vec{r} \phi, t \right) -p \left(\vec{r} \bullet, \vec{r}+\vec{b}_j \phi, t \right)
\right)
- k \left( \vec{r} \right) p \left(\vec{r}, t \right) , 
\label{eq:inhomo_1}
\end{align}
where $p \left(\vec{r} \bullet, \vec{r'} \phi, t \right)$ denotes the joint probability  at time $t$ that  
the site $\vec{r}$ is occupied by a reactive particle and the site $\vec{r'}$ is empty. 
We note the relation, \cite{Kutner81,Evans92,Tammaro95}
\begin{align}
p \left(\vec{r}  \bullet, \vec{r'} \phi, t \right)= p(\vec{r},t) -  p \left(\vec{r}  \bullet, \vec{r'} \circ, t \right)- p \left(\vec{r}  \bullet, \vec{r'} \bullet, t \right) , 
\label{eq:inhomo_2}
\end{align} 
where $p \left(\vec{r} \bullet, \vec{r'} \circ, t \right)$ and $p \left(\vec{r} \bullet, \vec{r'} \bullet, t \right)$ denote 
the joint probabilities at time $t$ that  
the site $\vec{r}$ is occupied by a reactive particle and the site $\vec{r'}$ is occupied by an inert particle 
and that both sites are occupied by reactive particles, respectively.
Equation (\ref{eq:inhomo_1}) can be rewritten as
\begin{align}
\frac{\partial}{\partial t} p \left(\vec{r}  , t \right)&= \gamma_B \sum_{j} \left( p \left(\vec{r}+\vec{b}_j  \bullet, t \right)
- p \left(\vec{r}  \bullet, t \right)
-p \left(\vec{r}+\vec{b}_j  \bullet, \vec{r} \circ, t \right)+ p \left(\vec{r} \bullet, \vec{r}+\vec{b}_j \circ, t \right)
\right) \nonumber \\
&- k \left( \vec{r} \right) p \left(\vec{r}, t \right) . 
\label{eq:inhomo_3}
\end{align}
By assuming that the pair correlation function depends on the distance vector between 
the reactant and the inert particle alone, 
the joint probability function can be written as, \cite{Moleko}
\begin{align}
p \left(\vec{r} \bullet, \vec{r'} \circ, t \right) = p \left(\vec{r}, t \right) c \left(\vec{r'}, t \right) \sigma(\vec{r}-\vec{r'}) , 
\label{eq:inhomo_4}
\end{align}
where the occupation probability by an inert particle is denoted by $c \left(\vec{r}, t \right)$. 
When the spatial variation of $\sigma(\vec{r})$ is smaller than $ p \left(\vec{r}, t \right)$ and 
$c \left(\vec{r}, t \right)$, 
we obtain in the limit of small lattice spacing, 
\begin{align}
\sum_{j} p \left(\vec{r}+\vec{b}_j \bullet, \vec{r} \circ, t \right) &= 
\left[ p \left(\vec{r}, t \right) c \left(\vec{r}, t \right) +
b^2 c \left(\vec{r}, t \right)\nabla^2 p \left(\vec{r}, t \right) 
\right]\sigma(0), 
\label{eq:inhomo_5_1}\\
\sum_{j} p \left(\vec{r} \bullet, \vec{r}+\vec{b}_j \circ, t \right) &= 
\left[ p \left(\vec{r}, t \right) c \left(\vec{r}, t \right) +
b^2 p \left(\vec{r}, t \right)\nabla^2 c \left(\vec{r}, t \right) 
\right]\sigma(0) , 
\label{eq:inhomo_5_2}
\end{align}
where $\lim_{b \rightarrow 0} (\vec{b}_j) =\sigma(0)$. 
By introducing Eqs. (\ref{eq:inhomo_5_1})-(\ref{eq:inhomo_5_2}), 
Eq. (\ref{eq:inhomo_3}) in the limit of $b \rightarrow 0$ becomes, 
\begin{align}
\frac{\partial}{\partial t} p \left(\vec{r}, t \right)  =  \vec{\nabla}\cdot D_B
\left[ (1-\sigma(0) c_v \left(\vec{r}, t \right)) \vec{\nabla} p \left(r, t \right) 
+ \sigma(0) p \left(r, t \right)\vec{\nabla} c_v \left(\vec{r}, t \right)    \right]
- k \left( r \right) p \left(r, t \right) , 
\label{eq:inhomo_6}
\end{align}
where $D_B=b^2 \gamma_B$ and 
the concentration of inert particles is denoted by $c_v(\vec{r},t)$ in the continuum limit. 
In the mean field approximation in which $\sigma(0)=1$, 
the derivation follows from that given previously. \cite{Evans92,Tammaro95} 
Equation (\ref{eq:inhomo_6}) can be expressed in terms the correlation factor as, 
\begin{align}
\frac{\partial}{\partial t} p \left(r, t \right)  =  \vec{\nabla}\cdot D_B
\left[ (1-c_v \left(r, t \right)) f_c \vec{\nabla} p \left(r, t \right) 
+ \sigma(0) p \left(r, t \right)\vec{\nabla} c_v \left(r, t \right)    \right]
- k \left( r \right) p \left(r, t \right) , 
\label{eq:inhomo_7}
\end{align}
where $\sigma(0)=f_c+(1-f_c)/c_v \left(r, t \right)$. 
$f_c$ is given by Eq. (\ref{suve26}), 
though, strictly speaking, the calculation of the correlation factor $f_c$ 
is restricted to the homogeneous distribution of inert particles. 
In the mean field approximation, we have $\sigma(0)=1$  and 
Eq. (\ref{eq:inhomo_6}) leads to Eq. (\ref{suv_cont_1}).


\begin{references}

\bibitem{Ellis}
R. J. Ellis and A. P.  Minton, {\it Nature} {\bf 425}, 27 (2003). 

\bibitem{Schnell}
S. Schnell and T. E. Turner, {\it Prog. Biophys. Mol. Biol.} {\bf 85}, 235 (2004). 

\bibitem{Minton}
A. P. Minton, {\it J. Biol. Chem.} {\bf 276}, 10577 (2001). 

\bibitem{Kapral}
C. Echever\'{i}a, K. Tucci and R. Kapral, 
{\it J. Phys.: Condens. Matter} {\bf 19}, 065146 (2007).

\bibitem{Zimmerman}
S. B. Zimmerman, and S. O. Trach, {\it J. Mol. Biol.} {\bf 222}, 599 (1991).

\bibitem{Zhou}
H.-X. Zhou, G. Rivas, and A. P. Minton, {\it Ann. Rev. Biophys.} {\bf  37}, 375 (2008).

\bibitem{Yethiraj}
J. S. Kim and A. Yethiraj, 
{\it Biophys. J.} {\bf 96}, 1333 (2009).

\bibitem{Nakazato} K. Nakazato and K. Kitahara,
{\it Prog. Theor. Phys.} {\bf 64}, 2261 (1980).

\bibitem{Suzuki} Y. Suzuki, K. Kitahara, Y. Fujitani, and S. Kinouchi,
{\it J. Phys. Soc. Jpn.} {\bf 71}, 2936 (2002).

\bibitem{Okamoto07} R. Okamoto and Y. Fujitani, 
{\it J. Phys. Soc. Jpn.} {\bf 74}, 2510 (2005).

\bibitem{vanBeijeren85}
H. van Beijeren and R. Kutner, {\it Phys. Rev. Lett.} {\bf 55}, 238  (1985).

\bibitem{SekiPRE}
K. Seki and M. Tachiya, {\it Phys. Rev. E} {\bf 80}, 041120 (2009).

\bibitem{SekiJCP}
K. Seki, M. Wojcik  and M. Tachiya, {\it J. Chem. Phys.} {\bf 134}, 094506 (2011).
%

\bibitem{Dong}
W. Dong, F. Baros, and J. C. Andre, {\it J. Chem. Phys.} {\bf 91}, 4643  (1989).

\bibitem{Dzubiella}
J. Dzubiella and J. A. McCammon, {\it J. Chem. Phys.} {\bf 122}, 184902 (2005).

\bibitem{Sun}
J. Sun and H. Weinstein, {\it J. Chem. Phys.} {\bf 127}, 155105 (2007).

\bibitem{Dorsaz}
N. Dorsaz, C. De Michele, F. Piazza, P. De Los Rios, and G. Foffi, {\it Phys. Rev. Lett.} {\bf 105}, 120601 (2010). 

\bibitem{Bhatia}
D. P. Bhatia, M. A. Prasad and D. Arora, {\it Phys. Rev. Lett. } {\bf 75}, 586 (1995).

\bibitem{Arora}
D. Arora, D. P. Bhatia and M. A. Prasad, {\it J. Stat. Phys.} {\bf 84}, 697 (1996).

\bibitem{Burlatsky}
S. F. Burlatsky, M. Moreau, G. Oshanin and A. Blumen, {\it  Phys. Rev. Lett. } {\bf 75}, 585 (1995).

\bibitem{Sokolov} I.M. Sokolov, R. Metzler, K. Pant, and M. C. Williams, {\it Phys. Rev. E} {\bf 72}, 041102 (2005).

\bibitem{Lee2000} 
J. Lee, J. Sung, and S. Lee, 
{\it J. Chem. Phys.} {\bf 113}, 8686 (2000).

\bibitem{Shin2003} 
J. Park, H. Kim, and K.J. Shin, 
{\it J. Chem. Phys.} {\bf 118}, 9697 (2003). 

\bibitem{Zumofen_EV}
G. Zumofen, A. Blumen, and J. Klafter, {\it Chem. Phys. Lett.} {\bf 117}, 340 (1985). 

\bibitem{Zhou91}
H.-X. Zhou and A. Szabo, {\it J. Chem. Phys.} {\bf 95}, 5948 (1991).

\bibitem{Schmit}
J. D. Schmit, E. Kamber, and J. Kondev, {\it Phys. Rev. Lett.} {\bf 102}, 218302 (2009). 

\bibitem{Doi} M. Doi,  
{\it J. Phys. A} {\bf 9}, 1479 (1976). 

\bibitem{Kotomin} V. Kuzovkov and E. Kotomin, {\it Rep. Prog. Phys.} {\bf 51}, 1479 (1988). 

\bibitem{Peliti} L. Peliti, {\it J. Physique} {\bf 46}, 1469 (1985). 

\bibitem{Hughes}
B. D. Hughes,
{\it Random Walks and Random Environments} vol. 1 (Clarendon Press, Oxford, 1995)
and references cited therein.

\bibitem{Bardeen}
J. Bardeen and C. Herring, {\it Imperfections in Nearly Perfect Crystals} (John Wiley \& Sons, Inc., New York, 1952), p. 261.

\bibitem{Tahir-Kheli} R. A. Tahir-Kheli and R. J. Elliott, {\it Phys. Rev. B} {\bf 27}, 844 (1983).

\bibitem{Zhdanov} 
V. P. Zhdanov, {\it Surf. Sci.} {\bf 194}, 1 (1988).

\bibitem{Evans92}
J. W. Evans, J. Chem. Phys. 97, 572 (1992). 

\bibitem{Tammaro95}
M. Tammaro, M. Sabella, and J. W. Evans, J. Chem. Phys. 103, 10277 (1995). 

\bibitem{Tammaro} 
 M. Tammaro and J. W. Evans, {\it J. Chem. Phys.} {\bf 108}, 762 (1998).
 
 \bibitem{Liu} 
 D-J. Liu and J. W. Evans, {\it J. Chem. Phys.} {\bf 113}, 10252 (2000)

\bibitem{McEwen} J.-S. McEwen, P. Gaspard, T. Visart de Bocarm\'{e} and N. Kruse,
{\it Surf. Sci.} {\bf 604}, 1353 (2010).

\bibitem{Tachiya78} M. Tachiya, {\it J. Chem. Phys.} {\bf 69}, 2375 (1978).

\bibitem{Sano} 
H. Sano and M. Tachiya, {\it J. Chem. Phys.} {\bf 71}, 1276 (1979).

\bibitem{Szabo} A. Szabo, K. Schulten and Z. Schulten {\it J. Chem. Phys.} {\bf 72} 4350 (1980). 

\bibitem{Collins} F.C. Collins, and G.E. Kimball, 
{\it J. Colloid. Sci.} {\bf 4}, 425 (1949).

\bibitem{Kutner81}
R. Kutner, {\it Phys. Lett. A} {\bf 81}, 239 (1981). 

 \bibitem{Moleko}
L. K. Moleko, Y. Okamura, A. R. Allnatt, {\it J. Chem. Phys.} {\bf 88}, 2706 (1988). 

\end{references}
\end{document}